%% file: astroph.tex
\documentclass[10pt,preprint]{aastex}

\usepackage{amsmath}
\usepackage{natbib}
\usepackage{epsfig}
\usepackage{graphicx}
\begin{document}
\bibliographystyle{apj}

\title{Combining Strong and Weak Gravitational Lensing in Abell~1689\altaffilmark{*}}

\author{
Marceau Limousin\altaffilmark{1},
Johan Richard\altaffilmark{2}, 
Eric Jullo\altaffilmark{3,4},
Jean-Paul Kneib\altaffilmark{4,2},
Bernard Fort\altaffilmark{5},
Genevi\`eve Soucail\altaffilmark{6},
\'Ard\'is El\'iasd\'ottir\altaffilmark{1}, 
Priyamvada Natarajan\altaffilmark{7,8},
Richard S. Ellis\altaffilmark{2},
Ian Smail\altaffilmark{9},
Oliver Czoske\altaffilmark{10},
Graham P. Smith\altaffilmark{11,2},
Patrick Hudelot\altaffilmark{12},
S\'ebastien Bardeau\altaffilmark{13},
Harald Ebeling\altaffilmark{14},
Eiichi Egami\altaffilmark{15}
\& Kirsten K. Knudsen\altaffilmark{16}
}

\altaffiltext{*}{Based on observations obtained at the Canada-France-Hawaii Telescope (\textsc{cfht}) which is operated by the National Research Council of Canada, the Institut National des Sciences de l'Univers of the Centre National de la Recherche Scientifique of France,  and the University of Hawaii.
Also based on observations from the \textsc{nasa/esa} \textit{Hubble Space Telescope} 
(Programs \# 9289 and 10150) obtained at the Space Telescope Science Institute, which is operated by \textsc{aura} under \textsc{nasa} contract \textsc{nas}5-26555
}
\altaffiltext{1}{Dark Cosmology Centre, Niels Bohr Institute, University of Copenhagen, Juliane Maries Vej 30, 2100 Copenhagen, Denmark; marceau@dark-cosmology.dk}
\altaffiltext{2}{Department of Astronomy, California Institute of Technology, 105-24, Pasadena, \textsc{ca} 91125, \textsc{usa}}
\altaffiltext{3}{European Southern Observatory, Alonso de Cordova, Santiago, Chile}
\altaffiltext{4}{\textsc{oamp}, Laboratoire d'Astrophysique de Marseille - \textsc{cnrs-umr} 6110 - Traverse du siphon, 13012 Marseille, France}
\altaffiltext{5}{Institut d'Astrophysique de Paris, Bvd Arago, 75014 Paris, France}
\altaffiltext{6}{Laboratoire d'Astrophysique de Toulouse-Tarbes, \textsc{cnrs-umr} 5572 \& Universit\'e Paul Sabatier Toulouse III, 14 Avenue Edouard Belin, 31400 Toulouse, France}
\altaffiltext{7}{Astronomy Department, Yale University, \textsc{p.o.} Box 208101, New Haven, \textsc{ct} 06520-8101, \textsc{usa}}
\altaffiltext{8}{Department of Physics, Yale University, \textsc{p.o.} Box 208101, New Haven, \textsc{ct} 06520-8101, \textsc{usa}}
\altaffiltext{9}{Institute for Computational Cosmology, Durham University, South Road, Durham \textsc{dh1} 3\textsc{le}, \textsc{uk}}
\altaffiltext{10}{Kapteyn Astronomical Institute, \textsc{p.o.} Box 800, 9700 \textsc{av} Groningen, Netherlands}
\altaffiltext{11}{School of Physics and Astronomy, University of Birmingham, Edgbaston, Birmingham, \textsc{b}15 2\textsc{tt}, England}
\altaffiltext{12}{Argelander-Institut f\"ur Astronomie, Universit\"at Bonn, Auf dem H\"ugel 71, 53121 Bonn, Germany}
\altaffiltext{13}{\textsc{l3ab} - \textsc{cnrs-umr} 5804 - 2, rue de l'Observatoire, \textsc{bp} 89, 33270 Floirac, France}
\altaffiltext{14}{Institute for Astronomy, University of Hawaii, 2680 Woodlawn Drive, Honolulu, \textsc{hi} 96822, \textsc{usa}}
\altaffiltext{15}{Steward Observatory, University of Arizona, 933 North Cherry Avenue, Tucson, \textsc{az} 85721, \textsc{usa}}
\altaffiltext{16}{Max-Planck-Institut für Astronomie, Königstuhl 17, Heidelberg, \textsc{d}-69117, Germany}

\begin{abstract}
We present a reconstruction of the mass distribution of galaxy cluster Abell~1689 at
$z = 0.18$ using detected strong lensing features from deep
\textsc{acs} observations and extensive ground based
spectroscopy. Earlier analyses have reported up to 32 multiply imaged
systems in this cluster, of which only 3 were spectroscopically
confirmed. In this work, we present a parametric strong lensing mass
reconstruction using 
34 multiply imaged systems of which 24 have newly determined
spectroscopic redshifts, which is a major step forward in building a robust
mass model.
In turn, the new spectroscopic data allows a more secure
identification of multiply imaged systems. 
The resultant mass model enables us to reliably predict
the redshifts of additional multiply imaged systems for
which no spectra are currently available, and to use the location of 
these systems to further constrain the mass model.
In particular, we have detected 5 strong galaxy-galaxy lensing systems just outside
the Einstein ring region, further constraining the mass profile.
Using our strong lensing mass model, we predict on larger scale a shear signal
which is consistent with that inferred from our large 
scale weak lensing analysis derived using \textsc{cfh12k} wide field images.
Thanks to a new method for
reliably selecting a well defined background lensed galaxy population,
we resolve the discrepancy found between the \textsc{nfw} concentration parameters 
derived from earlier strong and weak lensing analysis.
The derived parameters for the best fit \textsc{nfw}
profile is found to be $c_{200}=7.6\pm1.6$ and 
$r_{200}=2.16\pm0.10\,h^{-1}_{70}$ Mpc 
(corresponding to a 3\textsc{d} mass equal to 
M$_{200}= 1.32\pm0.2\times 10^{15}$ $h_{70}$ M$_{\sun}$).
We find that the projected mass enclosed within the Einstein
radius for Abell~1689 is 
M (45$\arcsec$) $ = 1.91\pm0.27\times 10^{14}$ $h_{70}$ M$_{\sun}$.
The large number of new constraints
incorporated in this work makes Abell~1689 the most reliably
reconstructed cluster to date. This well calibrated mass
model, which we here make publicly available, will enable us to exploit 
Abell~1689 efficiently as a gravitational
telescope, as well as to potentially constrain cosmology.
\end{abstract}


\section{Introduction}

Galaxy clusters are of considerable cosmological interest, as they are
the most recent structures to assemble in Cold Dark Matter
(\textsc{cdm}) scenarios with the largest angular scale. The
detailed distribution of mass in galaxy clusters offers a unique 
opportunity to test structure formation in the \textsc{$\Lambda$cdm} paradigm.
Cluster mass profiles can be probed via a variety of multi-wavelength
data - in the \textsc{x}-ray and optical, \textsc{sz} observations and
using measurements of the line of sight velocity dispersion of the cluster member galaxies.

The lensing effect, i.e. the bending of light by matter along the line
of sight from the source to the observer, depends only on the mass
distribution of the intervening structures, making gravitational
lensing an ideal tool for measuring the mass profiles of lensing
structures. In particular, no additional assumptions need to be made
with regard to the dynamical state (relaxed or not, in hydrostatic
equilibrium or not) or the nature (baryonic or not, luminous or dark)
of the intervening matter. However, \emph{all} mass distributions
along the line of sight contribute to the lensing signal, introducing
contamination by foreground or background objects. In the core of
massive clusters, the surface mass density is well above the critical
value enabling the use of detected strong lensing features to
constrain the inner part of the cluster potential. At larger
cluster-centric radii, the ellipticities of weakly sheared background
galaxies is used to estimate the weak lensing effects induced by the
cluster potential. 
Often one limitation of strong lensing is the
limited number of arcs and multiply imaged
systems available to probe the cluster potential. However, with deep
\emph{Hubble Space Telescope} (\textsc{hst}) exposure the situation is
changing and in this respect Abell~1689
is exceptional since it displays the largest number of multiple arc systems
ever identified at the center of a cluster.
 
In this work, we investigate the mass distribution of the
galaxy cluster Abell~1689, using both strong and weak lensing. This
galaxy cluster is one of the most \textsc{x}-ray
luminous $z\sim0.2$ Abell cluster, and has the largest Einstein radius,
around 45$\arcsec$, observed to date. Since
the size of the Einstein radius is related to the size of the critical
region of the cluster (i.e. the region where the surface mass density
is of the order of the critical surface mass density or larger)
this cluster was expected to exhibit many multiply imaged systems detectable in deep
\textsc{hst} exposures. While the field of view
of the Wide Field Planetary Camera 2 (\textsc{wfpc2})
was too limited to cover the full area of interest, with
the Advanced Camera for Surveys (\textsc{acs}), the situation has improved.
Indeed, it has provided an unprecedented wealth of arcs in
Abell~1689 \citep[][hereafter B05]{tom}, and many strong lensing studies
have been pursued by several groups taking advantage of this data-set.
Nevertheless, measurements of the mass distribution of Abell~1689
using various techniques were not satisfactorily convergent, particularly the
strong lensing and weak lensing analysis were not found quite compatible
in terms of the concentration parameter derived from fitting an \textsc{nfw} \citep{nfw} 
profile to the different data sets.

We use the \textsc{hst acs} data in 4 bands: F475W, F625W, F775W and
F850LP (see Fig.~\ref{nicefig}).
The composite color images are crucial to the identification of
multiple images, which must have the same colors, see also B05
for more details.
The new information we use in this work is the spectroscopic confirmation for
24 multiply imaged systems:
we have undertaken a long-term spectroscopic campaign using \textsc{keck}
and \textsc{vlt}, targeting singly and 
multiply-imaged sources in Abell~1689. Details of
the different observing runs are given in Richard et~al. (2007),
hereafter Paper I. While previous studies claimed 3 spectroscopically
confirmed multiply imaged systems to constrain their model (and many
others with photometric redshifts), we have expanded this to 24
spectroscopically confirmed multiply imaged systems. It is the first
time a galaxy cluster can be modeled using such a wealth of spectroscopic
information for arcs. We also use extensive ground based imaging:
to select cluster members, using
data from the \textsc{isaac} instrument on the \textsc{vlt} (see Paper I).
The region covered by \textsc{isaac} is a 3$\arcmin$ $\times$ 3$\arcmin$ square centered
on the \textsc{bcg}.

As an additional validation test of our mass model, we will compare the shear profile
predicted on large scales by the strong lensing model, and the shear
profile measured from \textsc{cfh12k} wide field data. The observation and data
reduction was done by \citet{olliphd} and the photometric
and lensing catalog was first constructed and described by \citet{bardeau05}.

Our first aim in this work is to build a reliable mass profile for
Abell~1689 and derive key structural parameters predicted by such as
mass concentration, existence of substructures, halo shape. 
Secondly, we use a new optimization procedure based on
the implementation of the Bayesian Monte Carlo Markov Chain method in
the \textsc{lenstool} software.
Thirdly, we present weak lensing measurements from \textsc{cfh12k}
data that match the predicted shear from our strong lensing
model. Hence, thanks to a better identification of background sources, we demonstrate 
reliably that both regimes are found to be in good agreement. Fourthly, we report the
finding of five strong galaxy-galaxy lenses located at $\sim
300$ kpc from the cluster center, i.e. \emph{outside} the critical
region of the cluster. Such a high rate of strong galaxy-galaxy
lensing events clearly points to the very high surface mass density of
this cluster.

This paper is organized as follows:
first we review in section~2 the results of previous modeling of
Abell~1689. Section 3 presents our
lensing methodology. In section 4, we describe the inclusion of the
multiply imaged systems.
Section 5 reports the new strong galaxy-galaxy lensing
systems we discovered. Our strong lensing results are presented in
section 6 and the weak lensing analysis in section 7. A discussion of
our results is presented in the concluding section, as well as a discussion
on the disagreement found in earlier works concerning the concentration
parameter of Abell~1689.

All our results are scaled to the flat, low matter density
\textsc{$\Lambda$cdm} cosmology with $\Omega_{\rm{M}} = 0.3, \
\Omega_\Lambda = 0.7$ and a Hubble constant \textsc{H}$_0 = 70$
km\,s$^{-1}$ Mpc$^{-1}$. In such a cosmology, at $z=0.18$, $1\arcsec$
corresponds to $3.035$ kpc. All the figures of the cluster are
aligned with \textsc{wcs} coordinates, i.e. North is up, East is left.
The reference center of our analysis is fixed at the \textsc{bcg} center:
\textsc{ra}=13:11:29.52, \textsc{dec}=-01:20:27.59 (J\,2000).
Magnitudes are given in the \textsc{ab} system.

\section{Previous Work}

\subsection{Optical Spectroscopy and Cluster Dynamics}

The spectroscopic study of the motions of cluster galaxies provides clues to the
dynamical state of the cluster.
It can reveal if the cluster is still undergoing a merger or if
it is already well relaxed \citep[see for instance work on the galaxy
cluster Cl0024+1654 by][]{olli00241,olli00242}.

In early work on Abell~1689, \citet{teague90} identified 176
cluster members, deriving a high velocity dispersion of $\sim$ 2\,355
km\,s$^{-1}$. 
This high value is somewhat misleading, as it reflects
the complex dynamical state of the cluster. This in fact emphasizes
the important role that lensing can play in the unbiased determination
of the mass distribution of clusters.
Subsequently \citet{girardi}, based on 96 redshifts in
the cluster center, divided the redshift distribution into three
distinct subgroups which overlap along the line of sight and reduced
the inferred central velocity dispersion down to $\sim$ 1\,429
km\,s$^{-1}$. 
Recently, \cite{lokas} used spectroscopic redshifts to study the
kinematics of about 200 galaxies in the cluster (mainly coming from the
\citet{teague90} work). They showed that the
cluster is probably surrounded by a few structures aligned along the
line of sight and concluded that the cluster mass cannot be reliably
estimated only from galaxy kinematics due to the complex structure in
Abell~1689. They also find that the inferred value of the velocity
dispersion depends sensitively on the choice of galaxy sample.
Due to technical improvements, in particular the advent
of high-multiplex spectrographs on 8-10 meter class telescopes, it is
now possible to get high-quality spectra for a large number of cluster
galaxies in a short time. The ongoing observations by
\citet{olliturin} using \textsc{vimos} on \textsc{vlt} report accurate
redshifts for 525 galaxies, spanning from the center outwards to 3
$h^{-1}$ Mpc, providing a deeper insight into the dynamics of this
cluster. Using the \citet{dressler88} test to probe for the presence
of substructure, \citet{olliturin} found only one apparent distinct
group of galaxies that lies $\sim$ 350 kpc to the north-east of the
cluster center. The location of this substructure corresponds to a
group of bright galaxies well identified in optical images of this cluster
(see Fig.~\ref{nicefig}). The redshift distribution of these galaxies
is skewed towards slightly higher redshifts. On larger scales (R $> 1\,
h^{-1}$ Mpc), no evidence for any substructures is found by the
\citet{dressler88} test; the outskirts of Abell~1689 look rather
homogeneous with no clear signature of any strong clustering.  The
fact that the large scale distribution is regular is also reflected in
the velocity dispersion profile which decreases from $\sim$ 2\,100
km\,s$^{-1}$ in the central bins to $\sim$ 1\,200 km\,s$^{-1}$ in the
outer bins (at a cluster-centric distance of $\sim$ 2\,$h^{-1}$ Mpc).
The very high velocity dispersion inferred for the central region
can be interpreted as a consequence of the non relaxed state of the
cluster inner regions, due to an ongoing merger.

To summarize, optical spectroscopy reveals that Abell~1689 is composed
of several structures aligned along the line of sight, and that a clear substructure is
present in the north-east.

\subsection{X-ray and SZ Studies}

The major assumptions made in most \textsc{x}-ray analysis, in order to
derive a mass profile from the observed surface brightness profile, 
is that the intra cluster gas is in hydrostatic
equilibrium in the cluster potential. This assumption is only
appropriate and valid for clusters that have had enough time to relax into
equilibrium after the last big merger event and that 
are not experiencing a merger event.

Abell~1689 has a rather circular \textsc{x}-ray surface brightness
distribution, which at first look suggests a relatively relaxed cluster. 
However, it has no cool core as often observed for relaxed clusters.
\citet{andersson}, then \citet{andersson2} presented a detailed measurement of the mass
from \textsc{x}-ray data obtained with \textsc{xmm} Newton. The temperature
map inferred from their analysis shows a clear discrepancy between the
northern and southern parts of the cluster, with a hint of a
temperature gradient in the southwest-northeast direction. The
redshift map shows a slightly higher-redshift structure to the east at
$z=0.185$, while the rest of the cluster is at $z\sim 0.17$. These
maps imply large-scale relative motion of the intra cluster gas, which
suggest that the cluster is likely not to be fully relaxed,
even if there is no bimodality in the \textsc{x}-ray emission.
Moreover, a lower than expected gas mass fraction also suggests a
complex spatial and dynamical structure.
To conclude, there is some evidence that Abell~1689 is on the late stage of
a merger, but this possible merger must be weak given that
departures of the gas from a relaxed state are modest.
It is common to see clusters with a quite regular shape in the \textsc{x}-ray map 
whereas the lensing mass map looks bimodal \citep{smith05}.

Furthermore, \citet{benson} measured the Sunyaev-Zeldovich effect in Abell~1689.
They found that the inferred optical depth of the comptonizing gas
might be higher than expected if one considers the simple spherically
symmetric model obeying hydrostatic equilibrium. They claim that an
elongated structure along the line of sight would alleviate the
discrepancy.

\subsection{Previous Weak Lensing Studies}

Abell~1689 is the first cluster in which a shear signal was
detected, both in the optical \citep{tyson90} as well as in infra-red
observations \citep{king02b}. It is also the first cluster for which
the lens magnification effect (depletion curve) has been used to
measure the absolute mass \citep{taylor98,dye01}.

\citet{tyson90} detected the weak lensing effect on a relatively small
field of view (720$\times$720 kpc$^2$), and selected the background
galaxy population by their blue \textsc{b}-\textsc{r} color. This early
work concluded that the inferred dark matter distribution correlates
well with the projected light distribution. Later on, \citet{tysonfisher95}
extended the weak shear analysis to $1\,h^{-1}$\,Mpc from the cluster
center. They found that the surface mass density follows a steeper
than isothermal profile on scales from $200\,h^{-1}$\,kpc to
$1\,h^{-1}$\,Mpc. Comparing with \textsc{n}-body simulations in
various cosmogonies, they found evidence for the $\Lambda > 0$ 
\textsc{cdm} model.

Wide Field observations of Abell~1689 were acquired on the
\textsc{eso/mpg} Imager \citep{clowe01,king02a}
and \textsc{cfh12k} camera at \textsc{cfht} \citep{bardeau05},
allowing to probe the cluster from the center to beyond the
virial radius. These three
studies used only \textsc{r} band imaging, selecting a background galaxy
catalog with a magnitude cut-off. The inferred shear profiles appear
flat in the inner region, likely due to contamination of their
background catalogs by cluster members. The Einstein radius was
estimated to be of the order of 22$\arcsec$ (much smaller
than the currently-accepted value of 45$\arcsec$). \citet{clowe03} and
\citet{bardeau06} later included \textsc{b} and \textsc{i} imaging from \textsc{cfht} to
better select a background galaxy catalog with a color cut-off. The
derived shear profile is then found to be steeper in the inner part,
indicating a weaker contamination by cluster members.
The M$_{200}$ mass is found to change little with the color selection,
but the \textsc{nfw} concentration parameter increases (see Table~\ref{cNFW}).
\citet{tomsubaru} from \textsc{subaru} multi-color data,
using a color cut-off to select background members, found a much
steeper shear profile than the one measured in all previous studies,
and a combined fit to this weak lensing data and the strong lensing data gives
an \textsc{nfw} concentration parameter of 10.8, 
which is not fully compatible with the concentration parameter derived from the
strong lensing studies.
\citet{elinor}, using similar \textsc{subaru} data, when fitting an \textsc{nfw} profile
to the weak lensing data \emph{only}, finds a concentration parameter as high
as 22.
Moreover, \citet{umetsu} used the same \textsc{subaru} multi-color data to infer a 2\textsc{d} mass
reconstruction from combined distortion and magnification data of their background
galaxy sample. This study confirmed the 1\textsc{d} analysis by \citet{tomsubaru}, in
particular the high concentration parameter.
Recently, \citet{1689flexion} measured the shape of the background galaxies present
in the \textsc{acs} field of view to estimate the reduced shear signal at a radius of
100$\arcsec$ from the center of the cluster. They find $g=0.2 \pm0.03$, a value which is
smaller than the one derived from \textsc{subaru} data by \citet{tomsubaru} and \citet{elinor} (who found $g\sim0.4$).
Note that the contamination by foreground members in this study is likely to be limited,
since the authors have a much better control of this issue than the ground based studies: they
analyze a small field of view where most of the cluster members have spectroscopic
measurements. Moreover they are going much deeper than ground based studies,
so the relative contamination is likely to be weaker.
This study identifies a second mass concentration located somewhere
in the north-east part of the \textsc{acs} field, but they were not able to accurately
constrain its location.

\subsection{Previous Strong Lensing Studies}

The first strong lensing modeling study of Abell~1689 was done by
\citet{miraldababul95}, using the detection of giant arcs from
ground based data by \citet{tyson90} to construct a simple mass model.
This early study already required two mass clumps, a central one and a second one
associated with the galaxy group in the north-east.
They pointed out a discrepancy, of a factor greater than 2, between the lensing and 
\textsc{x}-ray mass estimates, that is likely to arise from the existence
of these two clumps.

The new \textsc{acs} instrument has allowed strong lensing studies on
Abell~1689 to be pursued with greater accuracy.
The first published strong lensing study of
\textsc{acs} observations of Abell~1689 is the work by B05.
This remarkable analysis presented the discovery and
identification of 30 multiply imaged sources, spanning redshifts 
from $z\sim 1$ up to
5.5. Subsequent strong lensing works on Abell~1689 have benefited from 
this early identification, although some authors disagree on the identification
of a few multiply imaged systems. The modeling
technique employed by B05 is partly non parametric, and places the initial
large-scale mass clump aligned with the light distribution, but finds
a final mass model which is distributed rather differently from the
light distribution. \citet{diego} used a fully non parametric method to
reconstruct the mass using the same data set, where they do not assume
that the initial mass distribution follows the light.
Their results are in agreement with those from B05.

\citet{zekser} presented a model where they used the same set of
multiple images as B05. In this analysis, profiles are used to
define the initial form of the galaxy component in agreement with the
light, but the halo component is allowed to vary using the
\textsc{nfw} profile and perturbations added to it. These
perturbations are modeled as shapelets
\citep{refregier}, adding flexibility in the modeling and thus
accommodating a broad range of density profiles from cuspy, using a
\textsc{nfw} profile to various modifications of it.
Shapelets are used to add perturbations to the
deflection field (and not the mass, which make things more complex). 
Interestingly, the resulting description of the \citet{zekser} analysis 
is dominated by a shapelet,
representing $\sim 85\%$ of the halo mass, whereas the \textsc{nfw} represent only 
$\sim 15\%$. Thus the shapelet contribution to the modeling is more
than a perturbation since it dominates the halo budget,
suggesting that the central part of Abell~1689 is not well described 
by a \textsc{nfw} profile.
They further report some problems with a few multiply imaged systems identifications, 
and they construct models using subsets of multiple images, which considerably improves
the goodness of the fit (reducing the mean source plane scatter 
from 1.19$\arcsec$ to 0.74$\arcsec$).

Recently, \citet{halkola}, hereafter H06, presented the first
fully parametric mass reconstruction using the \textsc{acs} data. They use
two large scale dark matter clumps, one associated with the center of
the cluster, and the other with the main north-eastern substructure.
They report the misidentification for a few multiply imaged systems, 
and propose new ones.
One of the key features of this study is the careful modeling of the
galaxy member perturbation to the main potentials. They use a \citet{bbs}
profile (\textsc{bbs}) to
model cluster galaxies, parameterized by a central internal velocity dispersion
and a half mass radius. They carry out an analysis using the
observed fundamental plane in Abell~1689 to derive velocity
dispersions for 176 galaxies with \textsc{ab} magnitudes brighter than
22, and use two different scaling laws to relate the half mass radius
to the velocity dispersion. They obtain a strong lensing model that
is slightly superior to that of B05 (spatial resolution of \textsc{rms} 
of 2.7$\arcsec$ compared to
3.2$\arcsec$), and claim the difference is most likely a result of the
careful inclusion of the cluster galaxies. 
Moreover, it is worth noting that the \textsc{rms} of H06 can be regarded
as superior to previous studies because of the small number of free parameters used
in the modeling compared to non-parametric approaches.

In order to probe the cluster potential on all scales, both B05 and H06
used weak lensing data from \textsc{subaru} \citep{tomsubaru} in their analysis.
When fitting the \textsc{subaru} shear profile \emph{only} with an \textsc{nfw} profile, 
H06 finds a concentration parameter of 30.
According to their analysis, for the \textsc{nfw} profile,
the parameters obtained from strong lensing and weak lensing disagree
at the $\sim\,3 \sigma$ level. The high concentration of an \textsc{nfw} profile
fit to weak lensing data is incompatible with both the strong lensing results
presented by H06 and B05.

From the above descriptions, it appears
that this massive cluster is not fully relaxed.
Optical spectroscopy reveals several structures elongated along the line of sight and
a sub group of galaxies that lies
$\sim$ 350\,kpc to the north-east of the cluster center.
Parametric strong lensing studies all need a second mass clump around
this galaxy sub-group in their modeling, and the recent flexion study
by \citet{1689flexion} confirms this result.
The absence of cool core also supports a non fully relaxed state for
Abell~1689, though the \textsc{x}-ray analysis does not support any
bimodality as observed in more extreme cases as the bullet cluster \citep{andersson2}.
Moreover, the large strong lensing cross section observed for Abell~1689, with 
an Einstein radius of about 45$\arcsec$, also points to some merging processes acting
in this cluster that boost the strong lensing cross section \citep{torri,fedeli1,fedeli2}.
This likely non relaxed state is observed in 70\% of clusters at $z \sim 0.2$
\citep{smith05} and is not at odd with simulations.
The very unsatisfactory point is that so far no coherent model was found
combining strong and weak lensing, and more precisely the concentration parameter
derived from \textsc{subaru} weak lensing data
was so high compared to the one derived from strong lensing and to expectations 
of massive clusters that form in \textsc{n}-body simulations. Therefore we decided to
revisit the lensing modeling of Abell~1689, because such a cluster
with so many constraints is the best bet to
provide a complete coherence between all the modeling efforts of its
mass distribution.

\section{Lensing Methodology}
We describe below the strong lensing mass reconstruction that we have
used for Abell~1689. Schematically, we have use the 
parametric method \citep{stronglensing} as implemented in the
\textsc{lenstool} software which is publicly available\footnote{
\underline{\sl http://www.oamp.fr/cosmology/lenstool/}}.
In short we propose that the projected mass is made of two main dark matter
clumps that are very close to, or possibly in the process of, merging.
Then, we have introduced the deflecting mass of
individual galaxies using common scaling laws that statistically links the parameters of the 
galaxies to their luminosity.
We use the observational constraints (multiply imaged systems) to
optimize the parameters used to describe the mass distribution: this
is what we refer to as optimization procedure in the following sections.
An alternative method is the non-parametric approach that have been used
by other workers. In Appendix we discuss the main differences between parametric 
and non-parametric methods.

\subsection{Modeling the Dark Components}
\label{individualgalaxies}

We model the dark matter component separately on small
and large scales. The large scale model considers clumps that correspond
to the more extended cluster mass component, in contrast to smaller-scale clumps
that are associated with individual cluster galaxies. 

In this work we have modeled all dark matter halos present in the cluster
by a truncated Pseudo Isothermal Elliptical Mass
Distribution \citep[\textsc{piemd}, derived from][]{kassiola} scaled to their 
appropriate masses: from galaxy halos to dominant cluster halos.
Detailed properties of this mass profile can be found in
\citet{mypaperI}. The density distribution is given by:
\begin{equation}
\label{rhoPIEMD}
\rho(r)=\frac{\rho_0}{(1+r^2/r_{\mathrm{core}}^2)(1+r^2/r_{\mathrm{cut}}^2)}
\end{equation}
This mass profile is parameterized by a central density, $\rho_0$,
linked to the central velocity dispersion, $\sigma_0$, which is
related to the depth of the potential well. It is described using two
characteristic radii that define changes in the slope of the density
profile. In the inner region, the profile is described by a core with
central density $\rho_0$. The transition region
($r_{\mathrm{core}}<r<r_{\mathrm{cut}}$) is isothermal, with
$\rho\simeq r^{-2}$. In the outer parts, the density progressively
falls from $\rho\simeq r^{-2}$ to $\rho\simeq r^{-4}$, introducing a
cut-off. Recent work by
\citet{slacs3} has shown that early type galaxies are found to be
isothermal in their inner parts, with no significant evolution with
redshift up to $\sim 1$, supporting the use of an isothermal model to
describe early type galaxies. The \textsc{piemd} has successfully
been used to model galaxy clusters
\citep{stronglensing,smith05,covone,a68} as well as early type elliptical
galaxies \citep{Priya1,Priya2,Priya3,mypaperII}. Note that when using
this profile to model galaxy scale dark matter halos, we fix
$r_{\mathrm{core}}$ to be arbitrary equal to 0.15\,kpc, making it very similar to
the \textsc{bbs} profile. Then $r_{\mathrm{core}}$ is scaled with 
the luminosity as explained below.

Each clump, of either class, can be fully characterized using seven
parameters: the center position (\textsc{ra, dec}), the ellipticity $e$,
position angle $\theta$, and the parameters of the density profile,
$\sigma_0$, $r_{\mathrm{core}}$ and $r_{\mathrm{cut}}$.

We consider two large scale dark matter clumps for Abell~1689. The first clump is
associated with the center of the cluster, and the second clump is
associated with the north-east substructure. As a first guess, the
position is set to coincide with the luminous barycenter of each
structure, but during the modeling the tolerance limits on the position
parameters are large enough
to allow the center of the clumps to differ from the luminous
component.
For each clump, we fixed the value of the $r_{\mathrm{cut}}$ parameter
to reasonable values (Table~\ref{tableresults}): this parameter describes the properties of the
mass distribution on large scales, much larger than the radius over which the strong 
lensing constraints can be found. Thus in practice strong lensing cannot give any reliable
constraints on this parameter.
For the second north-east clump, we decided to limit its velocity dispersion to $\sim$ 500\,km\,s$^{-1}$
since no \textsc{x}-ray emission is detected from this region.
The total number of free parameters describing the large scale dark matter clumps
is equal to 12.

The parameters describing the dark matter halos associated with
three individual galaxies were found to play an essential role for reproducing multiply imaged systems
(systems 1, 2 and 6) because they perturb some arclets \citep[see \emph{e.g.}][]{arcsubstructure},
thus they are also allowed to slightly vary in the optimization procedure.
The positions will be chosen within $\sim$1 kpc from the visible center of the
associated galaxy, and other parameters are allowed to vary
between reasonable limits during the optimization procedure.
Then we include the other cluster galaxies in the optimization procedure by applying
empirical galaxy scaling relations based on the galaxy luminosity in order to
reduce the number of free parameters.
Most of the cluster members are elliptical, but we also consider a
few late-type galaxies for which we have measured a spectroscopic
redshift equal to the redshift of the cluster.
Note that we only consider the galaxies that are observed inside the
\textsc{hst acs} field of view.

To extract early-type
cluster galaxies, we plot the characteristic cluster red-sequences
(\textsc{r}-\textsc{k}) and (\textsc{b}-\textsc{r}) in two
color-magnitude diagrams and select the objects lying on both
red-sequences as cluster galaxies (Fig.~\ref{redseq}). This yields
258 early-type cluster galaxies down to \textsc{k}=23.4.
We also include 9 galaxies which
were not located within the red-sequence, but for which we measured a
spectroscopic redshift equal to the redshift of the cluster (i.e. for
$0.16<z<0.22$, blue points on Fig.~\ref{redseq}). This cluster
galaxy population is incorporated into the lens model as potentials
described by Eq.~\ref{rhoPIEMD}, with parameters scaled as a function
of luminosity:
\begin{equation}
r_{\mathrm{core}}=r^*_{\mathrm{core}} \left(\frac{L}{L^*}\right)^{1/2} \qquad \& \qquad
r_{\mathrm{cut}}=r^*_{\mathrm{cut}} \left(\frac{L}{L^*}\right)^{1/2} \qquad \& \qquad
\sigma_0=\sigma_0^* \left(\frac{L}{L^*}\right)^{1/4}
\end{equation}
The scaling relation for $\sigma_0$ assumes that mass traces light,
and its origin resides in the \citet{FJ} relation, that has been found
to be reliable for describing early-type cluster galaxies
\citep[e.g.,][]{wuyts,fritz}.  Since the mass $M$ scales as
$\sigma_0^2 \, r_{\mathrm{cut}}$, we have: $M \propto L$, assuming
that the mass-to-light ratio is constant for all cluster members (note
however that the mass-to-light ratio is not constant as a function of radius
as we are finding that $r_{\mathrm{cut}} > R_{\mathrm{effective}}$). For a
given L$^*$ luminosity, we will search for the values of $\sigma_0^*$
and $r^*_{\mathrm{cut}}$, the only two free parameters, that gives the best fit, while
$r^*_{\mathrm{core}}$ is fixed at 0.15\,kpc.
Note that there are other possible scaling relations for $r_{\mathrm{cut}}$,
that we do not investigate in this work.

The other parameters describing the small scale clumps associated with 
early-type galaxies are
set as follows: the center of the dark matter halo is assumed to be
the same as for the luminous component, and the ellipticity and
position angle of the mass is assumed to be the same as the ones of
the light. These assumptions are supported by the work by
\citet{slacs3}, who found that for a sample of early type lens
galaxies: (i) isophotal and isodensity contours trace each other, and
(ii) the position angle alignment between the stellar component and
the singular isothermal ellipsoid lens model are coincident.
However, we caution that \citet{slacs3} considered early-type
field galaxies in their studies, and we extrapolate their results to early
type galaxies residing in a very massive galaxy cluster. It is not
clear whether or not this extrapolation is legitimate, and at this
stage we cannot prove this assumption further. Galaxy-galaxy lensing studies
from the \textsc{combo}-17 group \citep{martina}, \textsc{sdss} \citep{mandelbaum}
and \textsc{rcs} \citep{hoekstra04}
also suggest that the ellipticity of the light and mass are well correlated.
But we caution once again that these galaxy-galaxy lensing studies were for
field galaxies.

\begin{figure}[h!]
\begin{center}
\caption{Color magnitude diagrams and the selection of cluster member galaxies.
The red-sequence selection is shown in the red boxes: all galaxies in this box are considered
to be early-type cluster galaxies. The red points correspond to rejected galaxies, 
as we measured a redshift different from the cluster redshift.
The blue points correspond to galaxies for which we measured a redshift equal to the
cluster redshift, but that were not identified in both red sequences.}
\label{redseq}
\end{center}
\end{figure}

\subsection{Source Plane Optimization}
We perform the optimization in the source plane, i.e. we optimize the
fit by mapping the positions of the resolved multiple images back to the source
plane and requiring them to have minimal scatter.

The motivation for optimizing in the source plane is that it is much faster
in terms of computing time than optimizing in the image plane.
We are aware of the likely bias of magnification when optimizing in the
source plane, but
we verify that it becomes equivalent to fit in the source or the image
plane when the number of multiply imaged systems increases (see H06).
Beside, for images located close the the critical lines, the
$\chi^2$ calculation in the image plane is sometime uncertain, making the
source plane inversion more secure and stable. Furthermore, the total
number of images that constitute a multiply imaged system may be
unknown. Some of them can be demagnified and not detected, making
image plane optimization delicate, whereas any observed image can be
assigned a source position in the source plane.

A $\chi^2$ estimator is constructed to quantify how well each trial
lens model fits the observational data. Considering $j$ multiply
imaged systems and $k$ critical curve constraints, we have:
\begin{equation}
\chi^2{=}\sum_j  \chi^2_{\mathrm{pos}} (j) {+} \sum_k \chi^2_{\mathrm{crit}} (k)
\end{equation}
The first term, a sum over $\chi^2_{\mathrm{pos}} (j)$, is constructed as follows:
given a model, for each multiply imaged system $j$, we compute the source location
$\vec{u}_i^S = (x_i^S, y_i^S)$ for each observed image $\vec{u}^I_i \, (1 \leq i \leq \textsc{n})$,
using the lens equation: $\vec{u}_i^S\,=\,\vec{u}_i^I\,-\,\vec{\nabla}\varphi(\vec{u}_i^I)$, where
$\varphi$ is the lens potential.
Then the barycenter $\vec{u}^B = (x^B, y^B)$ is constructed from the \textsc{n} sources, and we define:
\begin{equation}
\chi^2_{\rm pos} (j) {=} \frac{1}{\textsc{n}} \sum_{i{=}1}^{\textsc{n}}\frac{(x_i^S{-}x^B)^2{+}(y_i^S{-}y^B)^2}{\sigma_{\rm pos}^2}
\label{theory:chipos}
\end{equation}
To estimate $\sigma_{\rm pos}$, we assumed that the typical
uncertainty in measuring the position of any image is
$\sigma_I=0.2\arcsec$. This uncertainty is lensed back into the source
plane, using the amplification $\textsc{a}$: $\sigma_I^2 =
\textsc{a}\,\sigma_S^2$ and we take $\sigma_S$ as $\sigma_{\rm pos}$.

The second term, a sum over $\chi^2_{\mathrm{crit}} (k)$, measures how
well the locations of points on a critical line are reproduced by the
model. This stringent constraint can be used in cases where we know
with great confidence where the critical curve passes through. 
We define ($x_{\rm ct}^{\rm obs}$,
$y_{\rm ct}^{\rm obs}$) and ($x_{\rm ct}^{\rm mod}$, $y_{\rm ct}^{\rm
mod}$) as the observed and model critical line positions respectively
and construct $\chi^2_{\rm crit}$, where $\Delta x_{\rm crit}$ and
$\Delta y_{\rm crit}$ are the uncertainties in the location of the
critical lines (set to 0.3$\arcsec$):
\begin{equation}
\chi^2_{\rm crit}{=}\frac{(x_{\rm ct}^{\rm obs}{-}x_{\rm ct}^{\rm mod})^2{+}(y_{\rm ct}^{\rm obs}{-}y_{\rm ct}^{\rm mod})^2}{\Delta x_{\rm crit}^2{+}\Delta y_{\rm crit}^2}
\end{equation}
This additional constraint in critical line position has been used in two cases ($k=2$): 
for the radial pair in system 5, we imposed the location of the radial
critical line at $z=2.63$ to bisect this radial pair; and for system
12, we impose the location of the tangential critical line at $z=1.83$ to bisect the
tangential pair. In practice, we have
$\chi^2_{\rm pos} >> \chi^2_{\rm crit}$ and $\chi^2 \sim \chi^2_{\rm
pos}$.

For such a complex cluster of galaxies as Abell~1689, the
former parabolic optimization available in \textsc{lenstool} does not work 
well as it is too involving and time
consuming. Furthermore, \textsc{lenstool} provided no estimate of the
errors on the optimized parameters, unless dedicated investigation 
of space parameter around the best model is performed \citep{smith05}.
In order to measure errors and to
avoid local $\chi^2$ minima, we implemented the Bayesian Monte Carlo
Markov Chain (\textsc{mcmc}) package \textsc{bayesys} (Skilling,
2004) in \textsc{lenstool}.
According to a user defined model parameter space, the 
\textsc{mcmc} sampler draws randomly mock models and compute their
$\chi^2$. Progressively, it converges to the most likely 
parameter space and then it draws some realizations in this 
parameter space. We use these models
to do statistics, in particular to compute error bars
on the estimation of the parameters. 
With this new optimization method, \textsc{lenstool} returns samples of
points but also the \textsc{evidence} associated to the model we are
optimizing. The \textsc{evidence} characterizes our confidence in the
model we optimize. It works as Occam's Razor (MacKay 1991) : if a
simple model with few free parameters gives a good fit, a complex
model giving the same fit will have a lower \textsc{evidence}.
We describe the Bayesian \textsc{mcmc} method
in a dedicated publication \citep{mcmc}.

In the following, in order to quantify the goodness of a fit, we quote
in Table~\ref{multipletable} the \textsc{rms} both in the source plane and in the 
image plane.
Note that in this modeling we only consider the positions of the images or the critical
curves as constraints, i.e.  we do not include any constraints on the
flux in this analysis. In principle, any optimization process should
include flux constraints. However, the optimization of the positions
is dominant, and the work by \citet{zekser} has demonstrated that adding flux
constraints did not have a strong impact on the properties of the
final recovered mass model. However, some multiple images have shown
unexplained surface brightness anomalies, \emph{e.g.} system 12.
That should deserve a peculiar study in term of caustic perturbation at
very small scale.
Also a supernova could affect the relative brightness distributions confusing
the relative fluxes.

\begin{figure}[h!]
\epsscale{0.9}
\caption{Color image from F475W, F625W and F775W observations. North
is up, east is left. The substructure associated with the north-east
bright galaxy groups is clearly seen in this image.  Size of the field
of view is 160$\arcsec\, \times \, 160\arcsec$, corresponding to 485
kpc $\times$ 485 kpc. New spectroscopically confirmed multiply imaged
systems proposed in this work are shown (i.e. new systems not proposed
in previous works and for which
at least one of the image has been observed spectroscopically).
Note how system 31 shows an Einstein cross configuration.}
\label{nicefig}
\end{figure}
\begin{figure}
\epsscale{1.0}
\caption{Multiple images considered in this work, both candidates
(red squares) and spectroscopically confirmed (blue circles). The central
galaxies as well as one in the north-east galaxy group have been
subtracted for clarity.
Size of the field of view is 160$\arcsec\, \times \, 160\arcsec$,
corresponding to 485 kpc $\times$ 485 kpc.}
\label{11systemes}
\end{figure}
\include{table}

\section{Multiply Imaged Systems}

\subsection{Methodology}

In considering multiply imaged systems, we benefit from the work of
B05 as a starting point and we have followed their notation hereafter.
For details on the multiply imaged systems, see B05 or H06.
In Appendix we give more details and illustrations on the new spectroscopically
confirmed multiply imaged systems we consider. Since the redshift
of the images is crucial in constructing an accurate mass model, we
first considered systems for which we have measured an accurate
spectroscopic redshift for at least one of the multiple images. For each system, we
verified that the geometrical configuration was coherent and
checked the colors and morphology of the different members before
including a system in the optimization procedure. Our multiply imaged
system catalog in general agrees well with the one from B05. However,
we had to correct the identification for a few systems, and we propose
new multiply imaged systems. All multiple images included
in this work are shown on Fig.~\ref{11systemes}, and their coordinates
and redshifts are given in Table~\ref{multipletable}.
If the availability of a great number of redshift is a strong advantage it is still
challenging to detect reliably all the images of a given system,
specially the central ones that falls behind the central bright galaxies.
We preferred excluding rather than including the images for which
we are not sure of the identification.

The very first model was constructed using as constraints the images
constituting system 1 ($z=3.05$) and system 2 ($z=2.53$). Even at this
stage, a second large dark matter clump, associated with the galaxy
group at north-east, was needed to accurately reproduce the location
of the images of these two systems.
Then we included the other spectroscopically confirmed multiply imaged
systems. Before doing so, we use the current model to predict the location of the
different images that constitute this new system. To do so, we use
the image(s) for which we have measured a spectroscopic redshift as an
input. This is useful to check if any new multiply imaged system
agrees well or not with the current model. For example, this
procedure indicated that system 12, as proposed by B05, was problematic.

In order to better retrieve the locations of some images, we included 3 galaxies
explicitly in the optimization procedure (this means that the parameters 
describing these galaxies were
also considered as free parameters and optimized, as explained in
section~\ref{individualgalaxies}, instead of being scaled as a function of
luminosity). 
Indeed, these galaxies are located very close to some multiple images:
Galaxy 1 ($\alpha=197.859$, $\delta=-1.332$) is close to some images belonging
to systems 1 and 2. Galaxy 2 ($\alpha=197.886$, $\delta=-1.332$) is a bright
galaxy close to system 6. We also included the \textsc{bcg} to better retrieve
the central images and the radial system 5.
The inclusion of these extra free parameters was made possible since the 
number of constraints we have is large enough.
The parameters of these individual galaxies are presented in Table~\ref{tableresults},
keeping in mind that there may be a degeneracy between them and the 
cluster parametrization.
In total, we have 33 free parameters in the modeling.

\subsection{New Systems}

The new multiply imaged systems are shown on Fig.~\ref{nicefig}.
We propose five new spectroscopically confirmed multiply
imaged systems: system 32 at $z=3.0$ (Fig.~\ref{figimage32}); system 33 
at $z=4.5$
(Fig.~\ref{figimage33}); system 35 at $z=1.9$ (Fig.~\ref{figimage35});
system 36 at $z=3.0$ (Fig.~\ref{figimage36}) and system 40 at $z=2.5$ (Fig.~\ref{figimage40}).
We have split system 12 as proposed by B05 into two systems: systems 12
and 31 (Fig.~\ref{figimage31}). We began by following the work of B05, but did not
succeed in reproducing the 5 images they quote in their work with our
current model. By studying the individual images carefully, we found
that the morphology of the different images were inconsistent with
each other. We decided to restrict system 12 to two images, 12.2 and
12.3, forming a small highly amplified gravitational arc
which was already bisected by the critical
line at $z=1.83$ (the redshift of this arc). Moreover, object
12.1 in B05 is also spectroscopically confirmed at $z=1.83$, being part of
an other system, we renamed it 31.1. This object has a very characteristic shape,
constituted by two spots. The model predicted 4 counter images,
one of them being strongly demagnified in the central region of the cluster.
We were able to identify the three brighter counter images
with the correct relative orientation of the two little spots (Fig.~\ref{figimage31}).
We then constructed system 31 from these 4 images, giving an Einstein
cross like configuration (Fig.~\ref{nicefig}).
Note that H06 first proposed the splitting of system 12 from B05 into two
distinct systems.

We also provide a secure measurement for the redshift of system 10
($z=1.83$), that was used
in previous studies by B05 and H06 but with a wrong spectroscopic redshift 
estimation ($z=1.37$).
We do not consider systems 20 and 25 as proposed by B05 in our analysis.
For system 20, we expect a
third image which is not seen, and the morphology and colors of these
two images are somewhat different. Moreover, we took a spectrum for
the two images and found that if their redshift is similar, the
individual spectra are different (see Paper I for details).
Concerning system 25, we used image 25.2 from B05 to define a new system
that we call 33.

H06 also excluded system 20 from their analysis and proposed
the same identification as we do for system 12. Note that \citet{zekser} also
reported some problems with both systems 12 and 20, finding a
considerably improved goodness of fit when removing them from their
analysis.

\subsection{Candidate Systems}
At this stage of the modeling, we found that the mass of the second
large scale dark matter clump was not well constrained, as suggested
by the large error bars on the estimated parameters of the clump.
We interpreted this as due to the lack of observational constraints in this
region, as systems 6 and 30 are the only spectroscopically confirmed
systems in this region included in the
modeling. On the other hand, the position of the main dark matter
clump was found to be very well constrained. 

We used the current model to predict the
redshifts for all the other candidate multiply imaged systems. The
results are given in Table~\ref{multipletable}, where the errors bars
quote the 3$\sigma$ confidence levels inferred from the Bayesian
\textsc{mcmc} estimation. 
The estimates of the other candidate
multiply imaged systems are rather well constrained by the mass model 
and agree well with the photometrically estimated redshifts.
All the candidate systems are
assigned a redshift as estimated by the mass model. These systems are
then included in the optimization procedure as additional constraints.
We thus take benefit from all the available information to further
constrain the mass model. Note that even if the redshift of a multiply
imaged system is not known exactly, it is worth
using it in the optimization procedure since it will give us a net
positive number of additional constraints. For a system with known
redshift, and \textsc{n} multiple images, we get $2\,(\textsc{n}-1)$
constraints, but if the redshift is not known the number of
constraints reduces to $2\,(\textsc{n}-1) - 1$ which is positive as
$\textsc{n} > 1$.

\section{Strong Galaxy-Galaxy Lensing Events at Large Cluster-Centric Radii}

Inside the critical region ($R<100\arcsec$ from the center), multiple
images arise due to the cluster potential. Sometimes they appear close
to individual galaxies that locally boost the lensing signal. 

We also found some remarkable strong lensing features, located
\emph{outside} the critical region of the cluster (see
Fig.~\ref{stronggglensing}). Strong galaxy-galaxy lensing features
have been observed around many isolated galaxies, in particular around
early type galaxies in the field. They are still rare events, because
only a very favorable alignment between source, lens and observer can
produce these strong lensing events, due to the relatively small mass
of a single galaxy. The typical density of such events is
generally rare in galaxy field observations,
with about one strong galaxy-galaxy lensing event for about
1000 elliptical galaxies \citep{miralda}.
In this work, we detected five 
(of which three are clear lensing events)
strong galaxy-galaxy lensing events in a very small portion of the sky ($\sim 202^2 - \pi
\times 100^2$ sqr. arcsec. $\sim 2.8$ sqr. min., i.e. the size of the
entire field minus the size of the critical region of the cluster).
They would be very unlikely without the presence of the cluster potential,
which adds a strong external lensing signal ($\gamma \sim 0.22$ and
$\kappa \sim 0.36$ at 100$\arcsec$ from the center) to the lensing signal of the individual galaxies,
making strong lensing possible around these individual galaxies (see the
work by \citet{kovner} on these so-called 'marginal lenses'). These simple
observations suggest that this galaxy cluster is very massive.
It would be interesting to compare the density of strong galaxy-galaxy
lensing events that are found (or not found) for other clusters. This
density could be used in principle to diagnose the cluster's surface mass density in
this transition region between weak and strong lensing and bridge the
two lensing regimes.

These galaxy-galaxy lenses are of particular interest: with redshift
measurements, we could put direct lensing constraints on the mass
distribution of the lensing galaxies, stronger than the ones put on
the \textsc{bcg} and Galaxies 1 and 2. This is because the influence
of the cluster is not as strong as it is in the center of the cluster,
and therefore there will be less degeneracy between the cluster and
the galaxy parameterization. Moreover, with high resolution
spectroscopy, we could also combine lensing analysis with velocity
dispersion profiles as proposed by \citet{slacs3}, and test the hypothesis
we introduced before for modeling the cluster members (section 3.1). 
We list some properties of these galaxies in Table~\ref{gglens}.

\begin{figure}[h!]
\begin{center}
\epsscale{0.3}
\vspace{1cm}
\caption{Strong galaxy-galaxy lensing events detected outside the critical region of the cluster:
full color images constructed from F475W, F625W and F775W observations (inverse colors),
and images with the lens galaxy subtracted. Top row, from left to right: 
galaxies number 1, 2, 3. Bottom row, from left to right, galaxies number 4, 5.
Size of each panel is 30\,kpc$\times$30\,kpc.
The first three systems are clear strong lensing events, whereas the lensing interpretation is not as clear
for the two last events. Spectroscopic measurements are needed to alleviate the doubt.}

\label{stronggglensing}
\end{center}
\end{figure}

\begin{table*}
\begin{center}
\begin{tabular}[h!]{ccccccccc}
\hline
\hline
\noalign{\smallskip}
number & \textsc{ra} & \textsc{dec} & $\mathrm{d_{\textsc{bcg}}}$ & Lens $z_{\rm spec}$  & Lens magnitude & R$_{e}$\\
\noalign{\smallskip}
\hline
\noalign{\smallskip}
1  & 197.872 & -1.309 & 114 $\arcsec$ & 0.1758 & 18.25$\pm$0.13 & 2.0$\pm$0.1 \\
\noalign{\smallskip}
2  & 197.897 & -1.345 & 90 $\arcsec$ & 0.1844 & 18.58$\pm$0.16  & 2.2$\pm$0.2  \\
\noalign{\smallskip}
3 & 197.884 & -1.369 &  110 $\arcsec$ & 0.1855 & 18.11$\pm$0.01 & 4.1$\pm$0.1  \\
\noalign{\smallskip}
4 & 197.894 & -1.323 &  101 $\arcsec$ & -      & 18.69$\pm$0.15 & 2.6$\pm$0.2  \\
\noalign{\smallskip}
5 & 197.868 & -1.312 &  105 $\arcsec$ & 0.1990 & 18.64$\pm$0.09 & 3.0$\pm$0.2   \\
\noalign{\smallskip}
\hline
\hline
\end{tabular}
\caption{Strong galaxy-galaxy lensing events detected outside the critical region of the cluster: coordinates
(J2000), distance to the \textsc{bcg}; spectroscopic redshift measurements when available; F775W \textsc{ab} magnitudes obtained from
the surface brightness profile fitting by H06; circularized physical half light radius in units of kpc (H06).}
\label{gglens}
\end{center}
\end{table*}

\section{Mass Distribution from Strong Lensing}

The main advance we have made in the inner mass modeling compared
to previous works is to have included 34 multiply imaged systems
as constraints, of which 24 do have spectroscopic confirmation.
The positions of the observed images are reproduced accurately by our
model for the majority of the images.
The mean scatter in the source plane is equal to 0.45$\arcsec$.
This is much better than the mean scatter in the source plane 
reported by \citet{zekser} of 1.19$\arcsec$.
In the image plane, the \textsc{rms} is equal to 2.87$\arcsec$
This is comparable to previous studies performing the optimization
in the image plane: B05 reported a \textsc{rms} equal to 3.2$\arcsec$, and
H06 2.7$\arcsec$.
As a complementary confirmation of the quality of the lensing model,
we checked that all spectroscopically confirmed background sources for
which we did not identify multiply imaged systems were predicted to be
singly imaged by our model.

We begin this section by presenting the best fit \textsc{piemd}
parameters we get as an output of the optimization procedure\footnote{A
parameter file containing all the following information, and which can
be used with the publicly available \textsc{lenstool} software, is
available at {\underline{\sl http://www.dark-cosmology.dk/archive/A1689/}}. 
This file can be useful for
making model based predictions, e.g. counter-images of a multiple
image candidate, amplification and mass map and location of the
critical lines at a given redshift, and it will be updated. We also provide the mass map
generated from the best fit model as a \textsc{fits} file for direct
use.}.
In Appendix, we provide a comparison of the model presented here
with other strong lensing works. We also discuss the interpretation of our model and
the degeneracies we encountered.

\subsection{Parameters of the Dark Matter Halos}

The parameters found for the two large scale dark matter (\textsc{dm}) clumps are
given in Table~\ref{tableresults}.
The position of the main dark matter clump is found to be offset from the \textsc{bcg} by $\sim$ 25\,kpc
(Fig.~\ref{massmap}). This position is exactly where the lines connecting the two main radial arcs (systems 5 and 22)
bisects.
As we could have expected, this position is also where the line connecting images 31.1 and 31.3 and
the line connecting 31.2 and 31.4 bisect (remember system 31 displays an Einstein cross configuration).
The peak of the \emph{total} mass distribution (corresponding to the superposition of the main dark matter clump
and the \textsc{bcg})
however coincide with the \textsc{bcg} (green point on Fig.~\ref{massmap}).
Thus the peak of the total mass distribution is coincident with the \textsc{x}-ray center which is also found to 
coincide with the center of the \textsc{bcg} \citep{andersson2}.

The central mass distribution dominates the entire cluster,
as suggested by the centrally peaked \textsc{x}-ray emission.
However the second clump is also a substantial contributor
and is really required by the observational constraints. This
reinforces the idea that violent processes are ongoing in Abell~1689.

\begin{deluxetable}{cccccccc}
\tabletypesize{\scriptsize}
\tablecaption{
\textsc{piemd} parameters inferred for dark matter clumps considered in the optimization procedure.
Coordinates are given in arcseconds with respect to the \textsc{bcg}.
The ellipticity $e$ is the one of the mass distribution, expressed as $a^2-b^2/a^2+b^2$. Error bars correspond
to $1\sigma$ confidence level as inferred from the \textsc{mcmc} optimization.
Values into brackets are not optimized.
}
\tablehead{
\colhead{Clump} &
\colhead{\textsc{ra}} &
\colhead{\textsc{dec}} &
\colhead{$e$} &
\colhead{$\theta$} &
\colhead{r$_{\mathrm{core}}$ (\footnotesize{kpc})} &
\colhead{r$_{\mathrm{cut}}$ (\footnotesize{kpc})} &
\colhead{$\sigma_0$ (\footnotesize{km\,s$^{-1}$})}
}
\startdata
\noalign{\smallskip}
\noalign{\smallskip}
Clump 1 &  0.5 $\pm$ 0.2 & -8.5 $\pm$ 0.4  &  0.21 $\pm$0.01  &  90.4$\pm$1.0  &  99.4$\pm$3.9  & [1500]  & 1212.7$\pm$13.0 \\
\noalign{\smallskip}
\noalign{\smallskip}
\hline
\noalign{\smallskip}
\noalign{\smallskip}
Clump 2 & -70.7$\pm$1.4  & 49.1$\pm$3.4 &  0.7$\pm$0.03  &  78$\pm$2.5   &  66.6$\pm$6.0 &  [500]  & 542.8$\pm$3.6 \\
\noalign{\smallskip}
\noalign{\smallskip}
\hline
\noalign{\smallskip}
\noalign{\smallskip}
\textsc{bcg}    & -1.0$\pm$0.2 & 0.1$\pm$0.4 &  0.47$\pm$0.04  & 66.0$\pm$6.3 &   5.2$\pm$1.0 &  128.5$\pm$37.0  & 370.5$\pm$10.3 \\
\noalign{\smallskip}
\noalign{\smallskip}
\hline
\noalign{\smallskip}
\noalign{\smallskip}
Galaxy  1 & [49.0] & [31.5]  &  0.70$\pm$0.13  & 113.5$\pm$9.0  & 25.1$\pm$2.8  & 158.3$\pm$16.5 & 220.5$\pm$11.4 \\
\noalign{\smallskip}
\noalign{\smallskip}
\hline
\noalign{\smallskip}
\noalign{\smallskip}
Galaxy  2 & -45.6$\pm$0.5  & 31.6$\pm$0.8  &  0.77$\pm$0.04  & 45.5$\pm$2.3  & 17.2$\pm$2.2 & 179.5$\pm$7.5 & 357$\pm$20.2 \\
\noalign{\smallskip}
\noalign{\smallskip}
\hline
\noalign{\smallskip}
\noalign{\smallskip}
L$^*$ elliptical galaxy & --  & --  & --    &    --   &  [0.15] & 53.5$\pm$5.0  &  129.7$\pm$3.3 \\
\noalign{\smallskip}
\noalign{\smallskip}
\enddata
\label{tableresults}
\end{deluxetable}

We optimized over the parameters 
for three early-type cluster galaxies individually, rather than scaling their parameters 
with luminosity. This was necessary to explain observed multiple
image configurations in their vicinity, as they were found to
dominate the lensing signal locally. Thus, additional constraints were
derived for these individual galaxies. We present these galaxies, and
their best fit parameters in Table~\ref{tableresults}. One should
however bear in mind, that the parameter constraints for these
individual galaxies are likely somewhat degenerate with the parameters of the
large scale dark matter clumps.
This is particularly true for the clump we associated with the \textsc{bcg} and with Galaxy 2:
these values are too high to be representative of galaxies.

We included all the other identified cluster members, most of which were
elliptical galaxies, in the optimization. We scaled the parameters as
a function of luminosity using the scaling relation discussed in
section~\ref{individualgalaxies}. 
We considered a cluster galaxy with luminosity $L^*$ corresponding to a F775W magnitude of 17.54. 
At the end of the optimization procedure, we get the best parameters
describing this $L^*$ galaxy.
We find $\sigma_0^* = 129.7\pm3.3$ km\,s$^{-1}$ and $r_{\mathrm{cut}}^* = 53.5 \pm 5.0$ kpc 
(1 $\sigma$).
Note that there can still be some degeneracy between the mass we put in the large scale dark matter
clumps and the mass associated with the galaxy cluster members. Keeping that in mind,
we find galaxy halos to have a rather small spatial extent (compared to
field galaxy halos).
Thus we provide evidence for truncation of galaxy dark matter halos in high
density environments.
Even if we are aware of the possible degeneracy mentioned above, we are confident in these values
since they are in good qualitative agreement with independent weak galaxy-galaxy lensing results
and with the expectations from numerical simulations.

\subsection{The Cluster Mass Profile}

In Fig.~\ref{massmap}, we show the \textsc{r} band image of 
Abell~1689 along with the contours generated from the
projected mass map inferred from the best-fit model (red contours).
These contours show where the
projected mass is equal to 1.6, 2.4, 4.0 $\times$ $10^{10}$ M$_{\sun}$\,arcsec$^{-2}$.
This mass map is found to be in very good agreement with the one presented by \citet{zekser}.
By integrating this two dimensional mass map, we get the total
mass profile shown in Fig.~\ref{comparmassprofile}. 
The horizontal line at the top of Fig.~\ref{comparmassprofile} shows at what distance,
from the center of the cluster, the
multiple images used to constrain the model are located, hence where
the mass estimate is robust. In
Fig.~\ref{comparmassprofile}, we compare the mass profile derived in
this work with the one from B05 and H06, and
we find a very good agreement. As a result the inferred concentration parameter
we find by fitting the inner mass profile with a \textsc{nfw} profile
($c_{200} = 6.0 \pm 0.6$ at 3$\sigma$ confidence level) is in good agreement
with previous estimations.
This is encouraging since different studies have used different techniques and different
sets of multiply imaged systems.
Also shown is the mass profile derived by \citet{andersson}
from \textsc{x}-ray data.
The observed discrepancy in these
independent estimates will be discussed in section~\ref{massdiscussion}.

To quantify the accuracy of the mass map,
we considered 1\,000 realizations coming out from the 
\textsc{mcmc} optimization procedure, and constructed
a mass map for each one. The maps where then stacked and the mean and
the standard deviation (dispersion) were computed at each
point. Then this dispersion was converted in a percentage on the
accuracy of the mass measurement.
The accuracy of the mass reconstruction was found to be smaller than 15\% over the whole field.
A similar procedure was used to estimate the error bars of the mass profile presented in
Fig.~\ref{comparmassprofile}.

\begin{figure}[h!]
\epsscale{1.0}
\caption{\textsc{r} band image of Abell~1689.
Size of the field of view is 160$\arcsec\, \times \, 160\arcsec$, corresponding to 485
kpc $\times$ 485 kpc. The red contours show where the
projected mass density equals 1.6, 2.4, 4.0 $10^{10}$ M$_{\sun}$\,arcsec$^{-2}$.
The precision on the mass measurement, as inferred from the \textsc{mcmc} optimization,
is found to be smaller than 15\% over the whole field.
The green blob shows where the peak of the mass map is found.
The blue circle shows the location of the main dark matter clump (3\,$\sigma$ limits).
}
\label{massmap}
\end{figure}

\begin{figure}[h!]
\epsscale{0.5}
\caption{Total projected mass as a function of the aperture radius.
The shaded area corresponds to the mass profile derived from this work,
within the 1$\sigma$ errors as inferred from the \textsc{mcmc} optimization.
The horizontal line at the top of the plot 
shows where multiple images are
found, hence where the mass estimate is more robust. For comparison
we plot the mass profiles from H06 (solid line), B05 (dashed line)
and Andersson \& Madejski (2004) from an \textsc{x} ray
analysis (dotted line).
For clarity, some error bars are excluded.}
\label{comparmassprofile}
\end{figure}

\subsection{Critical Lines}

The critical curves are well defined and illustrated in
Fig.~\ref{criticalines} for a source redshift equal to 3. The outer line
corresponds to the tangential critical curve, whereas the inner line
corresponds to the radial critical curve. 

\begin{figure}[!]
\epsscale{1.0}
\caption{Critical lines for a source redshift equal to 3. The outer
line is the tangential critical line, the inner line is the radial
one. Size of the field of view is 160$\arcsec\, \times \, 160\arcsec$,
corresponding to 485 kpc $\times$ 485 kpc.}
\label{criticalines}
\end{figure}

Caustic curves correspond to the unlensed location of the critical
curves. For a given system, the multiple image configuration will
depend on the location of the source with respect to the
caustics. Fig.~\ref{caustics} shows the radial and tangential caustics
for a source redshift equal to 1.83 (redshift of system 12, 31 and 10)
and the reconstructed location of the different sources (only the
sources corresponding to spectroscopically confirmed multiply imaged
systems are shown). The tangential caustic curve crosses source
number 12, resulting in a highly amplified tangential arc in the image
plane.

It is interesting to look at the distribution of the strongly lensed
sources. From Fig.~\ref{caustics}, we clearly see a group of 3
galaxies at $z=1.83$ (sources 10, 12 and 31) that have been strongly lensed by the cluster.
It has a maximal spatial extension of 6.5$\arcsec$, which translates
into 55 kpc.
Moreover, the redshift differences between these three galaxies translate
into velocity differences smaller than 300 km\,s$^{-1}$.
This means that these 3 galaxies are likely to be bound and interacting.
Moreover, we have other candidate multiply imaged systems
(systems 15 and 18) whose estimated redshift is compatible with $z=1.83$.
Thus it is possible that we are observing a galaxy group strongly lensed by Abell~1689.
Interestingly, this galaxy group would stand in the so called 'redshift desert'.
In addition we detect two pairs of likely interacting galaxies at $z=2.63$
and $z=3.05$. The properties of these galaxies are described in Paper I.

\begin{figure}[h!]
\epsscale{0.5}
\caption{Radial and tangential caustics at $z=1.83$ and the location
of the reconstructed sources. The radial caustic is the one with a circular shape.}
\label{caustics}
\end{figure}

\subsection{Mass Associated with the Galaxies}
We want to compare the mass associated with the individual galaxies
($M_{\mathrm{galax}}$), to the total mass,
($M_{\mathrm{tot}}$), as a function of radius (Fig.~\ref{mgalax}). 
Inside the Einstein radius ($R_{\rm{E}}\sim 136$ kpc), 
we find $M_{\mathrm{galax}}$ = 13\% $M_{\mathrm{tot}}$,
whereas inside a radius of 364 kpc (roughly the \textsc{acs} field
of view), we find that the contribution of the galaxies to the total mass
is 7\%.
As $M_{\mathrm{galax}}$ is proportional to the galaxy luminosity, these means that
we have a smaller mass-to-light ratio in the central part of the cluster compared
to the outer region, clearly showing the effect of mass segregation in cluster center.
Similar trend was also observed in Cl0024 \citep{kneib03}.
We caution the reader about the possible degeneracy between the smooth dark matter
component and the galaxy component. Moreover, we did not investigate other models to describe
the galaxies. However, we are confident in the \textsc{piemd} results presented here for the galaxy component
since they agree with independent studies of galaxy-galaxy lensing in clusters.
\citet{zekser} also compared the relative contributions from the galaxies and the halo, and found the 
galaxy component to account for 15\% of the total mass within the Einstein radius.

\begin{figure}[h!]
\epsscale{0.5}
\caption{Contribution of the galaxy component to the total mass as a function of radius.
The vertical dotted line shows the location of the Einstein radius R$_{\textsc{E}}$
(for a source redshift equal to 3).}
\label{mgalax}
\end{figure}

\section{Mass Distribution from Weak Lensing}

As an additional test for our strong lensing model, we check the
agreement with the weak lensing analysis at large radius.
Previous works have found a discrepancy between the 
\textsc{nfw} concentration parameters derived from weak and strong lensing separately.

\subsection{Shear Profile Measured from CFH12k Data}

We directly measure the shear profile of Abell~1689 in wide
field \textsc{cfh12k} data. A shear profile is constructed from the
shapes of the lensed background population, so any contamination of
the background catalog by foreground members will dilute the shear
signal. Any weak lensing study will need to minimize this
contamination, although it is not always easy and often impossible to avoid
for single band observations. However, this dilution of the weak
lensing signal by cluster members can also be turned to our advantage
and used to derive the properties of the cluster population, in
particular the radial light profile \citep[see recent work by][]{elinor}.

Shear profiles for Abell~1689 have previously been measured using the
same data set. \citet{bardeau05} used a magnitude cut-off to select
background sources, while \citet{bardeau06} used a color selection
(defined by \textsc{r}-\textsc{i} $>$ 0.7, corresponding to the color
of the red-sequence members). The latter provided a less contaminated
background catalog than the former, but the shear profile was still
found to be shallow in the center, where the cluster contamination is
more prominent. If the virial mass derived from these different studies roughly agrees,
the estimation of the concentration parameter differ significantly. 
Below we will investigate in detail this contamination problem through several
rejection criterion.

\paragraph{Construction of the Background Catalog:}

We performed a Bayesian
photometric study to estimate a redshift probability distribution
($P_{\mathrm{bayes}}$) for each object, based on the \textsc{b, r, i}
photometry. The details of the method are
given in \citet{mypaperII}, where we showed that this photometric
redshift estimate was reliable for galaxies with 
$z > 0.5$. We then apply a criterion to determine whether a galaxy 
is a foreground or a background galaxy, setting the limit at $z=0.3$.  
In particular, we first compute the quantity :
\begin{center}
\begin{equation}
\chi_{z}=\frac{1}{\textsc{n}}\int_{z}^{+\infty} P_{\mathrm{bayes}}(z')dz'
\end{equation}
\end{center}
where $\textsc{n}=\int_0^{+\infty} P_{\mathrm{bayes}}(z')dz'$ is a
normalization factor. This quantity provides information on the
fraction of the probability distribution which is beyond $z$. In
practice, the upper bound of the integral is set to 5.

We search for the threshold value of $z$, which best discriminates the
cluster members, while still leaving enough background galaxies for a
weak lensing study. To do so, we use a sub-sample of 239 elliptical
cluster galaxies for which we have spectroscopic redshifts
\citep[data from][]{olliturin}. These
objects are located out to 340$\arcsec$ from the center of the
cluster, i.e. where the contamination is most likely to be strong. 
We found that if we impose $\chi_{0.4}>$ 60, we actually get rid of
most of the contamination by the cluster members (with only 4\%
contamination remaining), while at the same time keeping enough
objects to pursue a weak lensing study, with a background source
density of 12 galaxies/arcmin$^2$.

The spectroscopic subsample of red cluster members used to calibrate our rejection criterion
does not allow to test the removal of the faint blue cluster members which
are also a source of contamination at large radius from the cluster
center. Considering the \textsc{r}-\textsc{i} color of the objects
which fulfill the criterion $\chi_{0.4}>$ 60, the majority ($\sim
60\%$) are red objects, with \textsc{r}-\textsc{i} $>$ 0.7. This
means that our criterion does select blue objects that could be also
cluster members. To test this residual contamination,
we compared the shear profile presented in Fig.~\ref{shearprofile4}
with the one inferred from a catalog whose members fulfill both
$\chi_{0.4}>$ 60 \emph{and} \textsc{r}-\textsc{i} $>$ 0.7. We find
both shear profiles to be comparable, in particular in the inner
part. We are therefore confident that our selection criterion does not
include too many blue cluster members.

The final catalog used to construct the shear profile contains objects
for which $\chi_{0.4}>$ 60. Fig.~\ref{shearprofile4} gives our shear profile,
along with the shear profiles derived by \citet{bardeau05} and
\citet{bardeau06}. All shear profiles agree well beyond
$R\sim400\arcsec$, where the contamination by cluster members becomes
negligible. Closer to the center of the cluster, the differences
between the different shear profiles increase, suggesting that
previous analysis do not minimize the contamination.

\paragraph{Discussion:}
Although not perfect, the rejection criterion we propose in this work seems to exclude most
of the cluster members, as it gives an expected non-flat shear
profile, in agreement with the predictions from the strong lensing
model (see next subsection). Another advantage of this rejection method is that it allows
keeping the distant lensed objects whose colors are as red as the
cluster red sequence, whereas a color cut-off will reject these
objects. In conclusion, we are able to keep more background objects than when
applying a color-cut off (10\,280 instead of 6\,280), resulting in a shear
profile with higher signal to noise ratio.

The analysis by \citet{1689flexion}
provides an estimation of the tangential reduced shear of about 0.2$\pm$0.03 at a
distance of 100$\arcsec$, in good agreement with our measurements
(Fig.~\ref{shearprofile4}). This is encouraging for the rejection
criteria we propose, especially because the analysis by \citet{1689flexion} relies on the
\textsc{acs} data and is likely not to suffer from contamination problems as
can be the case with wide field ground based studies.

Nevertheless, this improved rejection criterion described here
keeps some limitations. From the spectroscopic sub-sample, we estimate
a 4\% remaining contamination, but this may be underestimated since
the spectroscopic sub-sample is not fully representative of the whole
cluster population. Moreover, our Bayesian photometric redshift
analysis is based on only three colors informations.
We have shown \citep{mypaperII} that the redshift estimation was correct, 
but ultimately this background galaxy population should be constructed from a 
more robust photometric determination, using many filters.

\subsection{Shear Profile Predicted by the Strong Lensing Model}

The strong lensing analysis shows that the cluster has two large scale
dark matter clumps which dominate the mass distribution of the cluster as a whole, and  no
significant clump is found outside the region probed by the
\textsc{acs} data \citep{bardeau05}. In addition, the \textsc{x}-ray
emission is peaked near the center of the mass, making us confident
that we have constrained most of the mass of the cluster. We should
therefore be able to provide reliable predictions for the properties
on large scales, although they should be strictly considered as lower
limits.

We consider a background population distributed in a plane at
$z\sim0.9$, which is the mean estimate of the Bayesian photometric
redshifts for the background catalog we constructed. Background
galaxies are randomly distributed on the plane, with a density of 40
galaxies per square arcminute and shapes are assigned by drawing the
ellipticity from a Gaussian distribution similar to the observed
\textsc{cfht} ellipticity distribution. We only consider the
intrinsic noise corresponding to the width of the ellipticity
distribution. In particular, we assume that the \textsc{psf} can be
perfectly estimated and subtracted and do not introduce any
additional observational noise. We calculate the deformations
produced by our best-fit mass model for this cluster on this
background galaxy population. The resulting catalog of lensed images
is then used to construct a shear profile out to large radius. This
shear profile is plotted as a solid line in Fig.~\ref{shearprofile4}
and is referred to as the predicted shear profile.

We are aware of the limitations of the predictions we are making
here: the strong lensing constraints span up to 85$\arcsec$ from the cluster
center, thus the predictions on larger scales will be model dependent.
The farther away from the center of the cluster, the less reliable
the predictions from the strong lensing model, and the only meaningful
comparison between the weak and strong lensing regime can be made in the
region of overlap between the data sets. Strictly speaking, only the three first
data points of the measured reduced shear profile on Fig.~\ref{shearprofile4} can be compared to the predictions
of the strong lensing model (i.e. up to 150$\arcsec$ from the cluster center).
Note that this inner region is precisely where the disagreement between shear 
measurements is the largest.
The method exposed in this subsection can be used to predict the level of the shear
signal which is expected and compare it to the measured weak lensing signal. This
can be useful to test for contamination in weak lensing catalogs.
Fig.~\ref{shearprofile4} shows that the shear profile predicted by the strong
lensing model is found to be in agreement with the shear profile
measured using weak lensing.

The shear profile measured by \citet{tomsubaru} using \textsc{subaru} data,
is shown in Fig.~\ref{comparshear},
and compared to the one presented in this work, based on \textsc{cfht} data.
We also show the curve of the best \textsc{nfw} fit to our weak lensing measurements.
This fit is found to be good.
Comparing \textsc{subaru} and \textsc{cfht} reduced shear profiles, we find marginal agreement.
In particular, in the central 150$\arcsec$, the \textsc{subaru} shear profile is higher than the
\textsc{cfht} one, reaching values as high as 0.7 in the very center. 
As a consequence, when fitting an \textsc{nfw} profile to the \textsc{subaru} shear profile only, high values for the concentration
parameter are derived: H06 finds $c_{200} \sim$ 30, and \citet{elinor} finds $c_{200} \sim$ 22 when using
a slightly different background galaxy catalog than \citet{tomsubaru}.
This will be discussed in the next subsection.

\begin{figure}[h!]
\epsscale{0.5}
\caption{Comparison between shear profiles constructed using different
rejection criteria to select background lensed sources: black squares
- corresponds to the Bayesian photometric redshift based selection;
red circles - a color selection, and blue crosses - a magnitude
cut-off is used (some of the errors bars omitted for clarity). 
The number of objects corresponding to each rejection criteria is given
in brackets. We find that all shear profiles agree well beyond $R\sim400\arcsec$.
The solid line corresponds to the strong lensing prediction, and the green filled triangle
comes from the study by Leonard et~al., 2007.
}
\label{shearprofile4}
\end{figure}

\begin{figure}[h!]
\epsscale{0.5}
\caption{
Comparison of our measured shear profile (black squares) with the
one measured from \textsc{subaru} data  presented in Broadhurst et~al. 2005b (red circles).
The dashed line corresponds to the best \textsc{nfw} fit to our
shear profile, and the green filled triangle
comes from the analysis by Leonard et~al., 2007.
}
\label{comparshear}
\end{figure}

\subsection{The NFW Concentration Parameter}
The different studies performed on Abell~1689 have given discrepant
values for the concentration parameter.
In our study, we find the strong lensing to agree well with the
weak lensing analysis and gives $c_{200} \sim 7$.

\begin{table*}
\begin{center}
\begin{tabular}[h!]{cccc}
\hline
\hline
\noalign{\smallskip}
$c_{200}$ & Method & Reference & Remark \\
\noalign{\smallskip}
\hline
\noalign{\smallskip}
\noalign{\smallskip}
6 & \textsc{wl} & \citet{clowe01} & magnitude cut-off \\
\noalign{\smallskip}
4.7 & \textsc{wl} & \citet{king02a} & same data as \citet{clowe01} \\
\noalign{\smallskip}
7.9 & \textsc{wl} & \citet{clowe03} &  color selection \\
\noalign{\smallskip}
$3.5^{+0.5}_{-0.3}$ & \textsc{wl} & \citet{bardeau05} & magnitude cut-off ($\sim$ 25\,000 objects)\\
\noalign{\smallskip}
$5.2\pm 0.3$ & \textsc{wl} & \citet{bardeau06} & color selection ($\sim$ 6\,300 objects)\\
\noalign{\smallskip}
$22.1^{+2.9}_{-4.7}$ & \textsc{wl} & \citet{elinor} & based on \textsc{subaru} data \\
\noalign{\smallskip}
\hline
\noalign{\smallskip}
$6.5^{+1.9}_{-1.6}$ & \textsc{sl} & B05 &  non-parametric method\\
\noalign{\smallskip}
$5.7^{+0.34}_{-0.5}$ & \textsc{sl} & \citet{zekser} &  \textsc{nfw} (15\%) + Shapelets (85\%) \\
\noalign{\smallskip}
$6\pm 0.5$  & \textsc{sl} & H06 &  parametric method\\
\noalign{\smallskip}
\hline
\noalign{\smallskip}
$10.8^{+1.2}_{-0.8}$ & \textsc{wl + sl} & \citet{tomsubaru} & \textsc{acs} + \textsc{subaru} data (fitting surface mass density) \\
\noalign{\smallskip}
$7.6^{+0.3}_{-0.5}$ & \textsc{wl + sl} & H06 & \textsc{acs} + \textsc{subaru} data (fitting shear profile) \\
\noalign{\smallskip}
\hline
\noalign{\smallskip}
$7.7^{+1.7}_{-2.6}$ & \textsc{x}-ray & \citet{andersson} &  hydrostatic equilibrium assumption\\
\noalign{\smallskip}
\hline
\noalign{\smallskip}
\noalign{\smallskip}
$7.6\pm 1.6$ (1$\sigma$)& \textsc{wl} & This work &  \textsc{bpz} selection ($\sim$ 10\,300 objects) \\
\noalign{\smallskip}
$6.0\pm 0.6$ (3$\sigma$)& \textsc{sl} & This work &  \\
\noalign{\smallskip}
\hline
\hline
\end{tabular}
\caption{Concentration parameters found in different studies.}
\label{cNFW}
\end{center}
\end{table*}

Fitting an \textsc{nfw} profile to the reduced shear profile of
Abell~1689 (Fig.~\ref{comparshear}), 
we find: $r_{200}=2.16\,\pm0.10$ Mpc and $c_{200}= 7.6
\pm 1.6$. Considering the background catalog constructed with a
magnitude cut-off, \citet{bardeau05} found: $r_{200}=1.9$ Mpc,
$c_{200}=3.5^{+0.5}_{-0.3}$, and considering the background catalog
constructed with a color selection, \citet{bardeau06} found:
$r_{200}=2.2$ Mpc, $c_{200}=5.2^{+0.3}_{-0.3}$. Comparing these
values, we see that $r_{200}$ (and hence $M_{200}$) is a reliable
estimator that does not depend strongly on the contamination of the
background catalog by cluster members. This shows that the mass
estimate from weak lensing is a robust quantity, even if they are
based on contaminated background catalogs (as is often the case, in
particular for studies with data in one band only). On the other
hand, the values inferred for the concentration parameter strongly
depend on the inner part of the shear profile which is more affected
by the contamination. As less contamination in the background catalog
means a steeper shear profile, it leads to a higher estimate of the
concentration parameter.

It is interesting to compare the concentration parameters found in
other studies since the values for $r_{200}$ roughly agree with each
other (Table~\ref{cNFW}). \citet{clowe01}, considering a background catalog
constructed with a magnitude cut-off, found: $c_{200}\sim 6$, and when
using a more refined maximum likelihood method to fit the data,
$c_{200}=4.7$ was found \citep{king02a}. \citet{clowe03}, using color
information from \textsc{cfht} (the same as the one used in this
paper) to better select background sources, and using a maximum
likelihood method, found $c_{200} \sim 7.9$. \citet{tomsubaru} found a higher
concentration, $c_{\mathrm{vir}}\sim 13$ (corresponding to $c_{200}$=
10.8) when fitting \emph{simultaneously} both strong and weak lensing data points.
However, H06, when fitting their strong lensing based mass profile
with the weak lensing \textsc{subaru} data, found $c_{200}=7.6^{+0.3}_{-0.5}$, when \citet{tomsubaru}
was finding $c_{200} \sim 10.8$. This is surprising, because the inner mass profiles
inferred by B05 and H06 from strong lensing agree well, and both studies use the same weak lensing
data from \textsc{subaru}. The only difference between these two fitting is that
B05 fit the \textsc{nfw} profile to the surface mass density map obtained
from the shear values, whereas H06 fit the shear directly. But this is not likely
to explain such a large difference, since both quantities are two different ways of
characterizing the same mass distribution. The only explanation may come from the
relative weights that have been assigned to the weak lensing and strong lensing
data points when constructing an overall $\chi^2$
since both regimes are found in disagreement by B05 and H06. Assigning different 
weights to the different data points will lead the $\chi^2$ to 
be more representative of the strong lensing or the weak lensing data points.
Another contribution may come from the different errors estimations in the strong lensing
mass profile found by B05 and H06.
This demonstrates that if the weak lensing and strong lensing regime do not match
each other, it is not possible to give a reliable estimation of the concentration
parameter.
In that sense, the value of 7.6 found by H06 is misleading because 
this provides a bad fit to the \textsc{subaru} weak lensing data (see Fig.~21 of H06 and compare with our Fig.~\ref{comparshear}),
over predicting the outer radial profile. As a consequence, they find $r_{200}=2.55$ Mpc, which is larger than the value found in this work,
and which leads to a M$_{200}$ estimate which is twice the one presented here.
In this work the fact that the strong lensing profile and the weak lensing
profile are fully compatible offer the best opportunity to derive a good estimation
of the concentration parameter.

The concentration parameters derived from our weak lensing analysis 
($c_{200}= 7.6\pm 1.6$) agrees with the one derived from our strong lensing
analysis ($c_{200}= 6.0\pm 0.6$). Thus we reconcile the concentration parameters
derived from the two lensing regimes.
This values also agree with the one inferred from \textsc{x}-ray data,
where \citet{andersson} found $c_{200}=7.7^{+1.7}_{-2.6}$.

For the time being, the weak lensing studies based on wide field multi-color observations
are giving very different results for the concentration parameter.
Fitting an \textsc{nfw} profile to the \textsc{subaru}
weak lensing data \emph{only} from \citet{tomsubaru} gives $c_{200} \sim 30$ (H06),
or $c_{200}=22.1^{+2.9}_{-4.7}$ \citep{elinor}, whereas fitting
an \textsc{nfw} profile to the \textsc{cfht} data presented here gives $c_{200}=7.6\pm1.6$.
The main difference between \textsc{subaru} and \textsc{cfht} reduced shear profiles
is in the central part (Fig.~\ref{comparshear}): the \textsc{subaru} reduced shear profile is higher, 
leading to higher concentration parameter than the \textsc{cfht} reduced shear profile.
Discriminating between high ($c_{200}>$ 20) and low ($c_{200}<$ 10) values of the concentration
parameter is of interest since a high value of the concentration parameter, if confirmed independently, 
will have important consequences for cosmological models. Thus the weak lensing issue on Abell~1689
is not settled yet, and an independent study is needed to discriminate between 
\textsc{subaru} and \textsc{cfht} shear profiles. Recently, an independent study provided \emph{one} data point: 
\citet{1689flexion} measured the reduced shear
to be equal to 0.2$\pm0.03$ at a distance of 100$\arcsec$ from the cluster center (Fig.~\ref{comparshear}), 
in agreement with the \textsc{cfht} data.
This estimation is likely to be not contaminated by cluster members since
the analysis relies on a small and deep \textsc{acs} frame where the authors
can control much better (from visual inspection and from the extensive
spectroscopy available for central cluster members) the eventual contamination by foreground members
compared to wide field weak lensing studies dealing with large catalogs
of galaxies. Moreover, the deepness of the \textsc{acs} data allow
\citet{1689flexion} to reach a density of background sources equal to $\sim$ 200
sources/arcmin$^2$.
This unique data point can help to resolve the discrepancy between \textsc{subaru} and
\textsc{cfht} shear profiles
reduced shears values at a distance of 100$\arcsec$ from the center of the
cluster. However, this data point is not enough to fully settle the point and
another independent study providing a measurement of the reduced shear profile on a larger
range of radius is needed.

\section{Discussion and Conclusions}
\label{massdiscussion}

Fig.~\ref{comparmassprofile} shows the excellent agreement
found by different strong lensing studies for the inner mass profile, though
they use different methods.
One of the main conclusion of this work is that a good agreement for the
concentration parameter is found by the different methods used (strong
and weak lensing and \textsc{x}-ray).
We can see from Table~\ref{cNFW} that the mean value of $c_{200}$ is around 7.
Note that the \textsc{x}-ray measurement appears to be quite efficient to
determine the concentration parameter. This is not surprising since so far it seems
that this is the best way to determine the universal profile of dark matter halos,
even providing one of the best proof of this universal profile.

This value ($c_{200} \sim 7$) is slightly higher than the average value predicted
by \textsc{n}-body simulations of cluster formation and evolution in the
$\Lambda$\textsc{cdm} cosmogony by \citet{bullock} that predicts a concentration
parameter $c_{200} \sim 5.5$.
It is possible to reconcile observed high values of the concentration parameter
with the predicted one, \emph{e.g.} invoking
triaxiality of the dark matter halo \citep{clowe04,oguri,gavazzi05,corless}, neighboring
massive structures \citep{king_corless}, or baryonic physics 
\citep[which leads to more concentrated dark matter halo profiles as gas cools in their
inner region, see \emph{e.g.}][]{gnedin}.
Indeed, none of the
earlier simulations by \citet{bullock} contain a statistically significant sample
of massive clusters that could likely produce the plethora of
strong lensing phenomena displayed by clusters like Abell~1689.
The Millenium simulation \citep{millenium}, with a large box size (500 $h^{-1}$ Mpc)
on a side would offer an appropriate comparison sample.
What is needed at the present time is a distribution of best-fit
concentration parameters for observed lensing clusters. Given the
large diversity in observed massive, lensing clusters an adequate 
comparison can be made with simulations only with a large sample
(20 clusters or so).

This apparent disagreement between observed and predicted concentration parameters
may not be interpreted as a sign of failure of the \textsc{cdm} paradigm.
\citet{jing} has shown that the quality of the \textsc{nfw} fitting depends on
whether the halo is in equilibrium, and that substructures degrade the fitting 
quality, because the \textsc{nfw} profile was found for equilibrium halos.
Keeping in mind that Abell~1689 is composed by two massive large scale dark 
matter halos, this means that we have to discuss the \textsc{nfw} fitting
parameters with care. The work by \citet{zekser} also points out that the
\textsc{nfw} profile may not be a good description of the central part
of Abell~1689. 
Recently, \citet{saha} deprojected the mass map of Abell~1689 and obtained
inner profiles consistent with $\rho \sim r^{-1}$, supporting the \textsc{nfw}
predictions. However, they deprojected the mass map assuming spherical
symmetry, which is a strong prior that may bias the measurement.

It is worth noting a remaining discrepancy between the mass
estimates from lensing and \textsc{x}-ray measurements. In general,
\textsc{x}-ray estimates agree well with gravitational lensing
estimates for clusters with a high concentration of central
\textsc{x}-ray emission (i.e. relaxed cooling flow clusters), but are
seemingly in disagreement for less centrally peaked clusters
\citep{allen}. Abell~1689, as discussed before, 
seems not to be fully relaxed and does not present any cool core, though the \textsc{x}-ray
emission is clearly unimodal.
\citet{rasia} studied a set of five galaxy clusters, resolved at high
resolution, in a hydrodynamic simulation, examining the systematics
affecting the \textsc{x}-ray mass estimates. They showed that for a
cluster undergoing a merger, the assumption of hydro-dynamical
equilibrium led to the mass being underestimated by 30\%, and that a
$\beta$ model gave even more discrepant results. However, a 30\%
correction would still not be enough to reconcile the \textsc{x}-ray
mass estimate with that from lensing for Abell~1689, but the discrepancy 
then becomes smaller, of the order of 40\%.
This remaining discrepancy may be understood if Abell~1689
is undergoing a merger along the line of sight, which remains a possible
scenario at this stage as suggested by the broad redshift distribution
of the cluster members and the \textsc{sz} study.
\cite{andersson} argue that if Abell~1689 was undergoing a merger
along the line of sight, the \textsc{x}-ray analysis would
underestimate the total mass by a factor of 2. 
The simple model (two perfectly spherical clusters aligned exactly along
the line of sight) used to derive the estimate is implausible, with a
more realistic model leading to a smaller correction factor.
Thus we need to better understand the three dimensional distribution of Abell~1689
to quantify accurately the underestimation of the \textsc{x}-ray analysis
coming from a merger along the line of sight.
Also it is not clear yet why lensing analyses need an extra mass component in the 
north-east (larger than the mass associated with the individual galaxies), whereas no \textsc{x}-ray
emission is detected from this region.

The lensing study of Abell~1689 is not fully satisfactory yet.
On the weak lensing part, significant discrepancies between independent measurements
of the shear profile do remain (Section~7.3) and concentration parameters larger than 20 have
been claimed, which if confirmed independently will have important and interesting
implications for cosmological models.
On the strong lensing part, the very satisfactory point is that independent studies
that used different methods and different catalogs of multiply imaged systems find very good
agreement for the inner mass profile (Fig.~\ref{comparmassprofile}).
However, parametric strong lensing studies (H06 and this work) do not give a satisfactory description of the
needed second dark matter clump in the north-east (whose presence has been recently confirmed
by \citet{1689flexion} and \citet{saha07}). Both studies experienced some degeneracies with this second clump (see
discussion in Appendix). Therefore we believe that
the \emph{needed} mass component in that region \emph{cannot} be well described with a single mass clump.
Another evidence for this statement is that the individual \textsc{rms} of the multiply imaged systems
observed in that region (systems 6, 13, 30 and 36) are larger than the total \textsc{rms}, especially
for system 30.
The problem may come from the interpretation of the strong lensing mass model.
One possible explanation could be that this second clump is a projection effect of a large
merging filament that appears in numerical simulations,
as suggested by the optical spectroscopy of the bright galaxies in the \textsc{acs} field.
Clearly we need to put some efforts into probing the three dimensional structure of Abell~1689,
in particular a lot could be learned from an extensive study of the line of sight velocity dispersion 
of the cluster members and from \textsc{sz} analysis.

In this paper, we have made the well established assumption that dark matter exists. 
However, modified theories of gravity are not ruled out by this work, 
and we encourage people working 
in that area, to make use of the extensive data set provided in this paper, to test their theories.

To summarize this work, we have presented an accurate mass model for Abell~1689, constructed
from 24 spectroscopically confirmed multiply imaged systems. This
makes by far the most strongly constrained cluster to date, in terms of
the number of multiply imaged systems included with spectroscopic identification.
Our results are in agreement with a previous study by
\citet{halkola}, as well as with the non parametric studies by
\citet{tom} and \citet{diego}, but are much more reliable given the
data we use. Additional spectroscopic observations are ongoing to
refine the model further.
We have performed a parallel weak lensing study, using wide field
images from the \textsc{cfh12k} camera and devised a new method to
select lensed background galaxies. We find very good agreement between
the strong and the weak lensing regimes, resolving the discrepancy
found in earlier work on this cluster, 
in particular we do not infer any high value for the \textsc{nfw} parameter
. Another important result of the weak lensing
analysis is that the weak lensing based mass estimates for galaxy
clusters are reliable, even when using one band imaging with
contaminated catalogs. With the goal of bridging the gap between the
strong and weak lensing regimes, we suggest a new method of diagnosing
the surface mass density in the intermediate region, by using the
density of detected strong galaxy-galaxy lensing events that we report here.
A detailed description of this new method will be described in a following paper.

The accurate mass model presented in this work is made publicly available,
thus we provide a well understood gravitational telescope to the community
and hope it will be used to go beyond the current observational facilities.

\acknowledgments

We thank the referee for a careful reading and a constructive report that improved the work presented here.
The Dark Cosmology Centre is funded by the Danish National Research Foundation.
This work is based on observations coming from the following facilities:
\textsc{hst acs}, \textsc{lris} on \textsc{keck}, \textsc{isaac} 
and \textsc{fors} on \textsc{vlt}, \textsc{cfh12k} on \textsc{cfht} (see Paper I).
ML thanks those people for useful discussion and
constructive comments, in particular: Cecile Faure, Douglas Clowe,
H{\aa}kon Dahle, Jens Hjorth, Karl Andersson,
Kristian Pedersen, Leon Koopmans, Phil Marshall, Tom Broadhurst.
ML also thanks Aleksi Halkola for many useful interactions during the
writing of this paper.
We thank John Skilling for allowing us to use his \textsc{bayesys
mcmc} sampler. We thank the Danish Centre for Scientific Computing at
the University of Copenhagen for providing generous amount of time on
its supercomputing facility. JPK acknowledge support from CNRS.
Argelander-Institut f\"ur Astronomie is founded by merging of the
Institut f\"ur Astrophysik und Extraterrestrische Forschung, the
Sternwarte, and the Radioastronomisches Institut der Universit\"at Bonn.
IRS and GPS acknowledge support from the Royal Society. This
work was supported by the European Community's Sixth Framework Marie
Curie Research Training Network Programme, Contract
No. MRTN-CT-2004-505183 "ANGLES".

\clearpage
\bibliography{astroph}

\section*{Appendix}
\section*{A - Comparison of our Best-Fit Model with other Strong Lensing Work}

As shown in Fig.~\ref{comparmassprofile}, we find a very good
agreement with inner mass profiles derived in previous studies by B05 and
H06. All studies also agree on the inferred concentration
parameter, $c_{200}\sim6$.

\subsection*{Comparison with Non-Parametric Methods}

Non-parametric methods are feasible when the number of constraints
becomes large enough, as is the case with the new \textsc{acs}
observations of Abell~1689. A clear advantage provided by
non-parametric methods is the flexibility, i.e. the allowed range for
the mass distribution is much wider than with parametric methods.
Abell~1689 has been successfully modeled using \textsc{acs} data and a
non parametric method by B05 and \citet{diego}. However, these
non-parametric works were able to reproduce accurately three sets of
unlikely multiple images (system 12, 20 and 25 of B05). Using a
parametric mass reconstruction, we were not able to accurately
reproduce these multiply imaged systems, and the parametric strong
lensing analysis by H06 led to the same conclusion. The large freedom
allowed by non-parametric methods, could result in the possibility of
a good reproduction of partly wrong data, without a clear possibility
to predict misidentifications of multiple images. However, the
important and encouraging point is, that the overall mass profiles
derived from parametric and non-parametric methods are in very good
agreement.

\subsection*{Parameterization of the Dark Matter Distribution}

\subsubsection*{On Large Scales:}

The parameters found for the two galaxy cluster scale dark matter
clumps roughly agree with the ones found by H06.
However, the description of the second north-east mass clump is not satisfactory yet.
There is good evidence from \textsc{x}-ray and lensing that the dominant mass distribution in
Abell~1689 is centrally distributed.
But some extra mass is clearly required in the north-east to reproduce the observations:
all parametric strong lensing studies need a second clump to reproduce the multiple 
images configuration, even the very first analysis by \citet{miraldababul95} which is based
on ground based imaging, and recent
analyses by \citet{1689flexion} and \citet{saha07} also provide evidence
for substructure in that region.

During the modeling of Abell~1689, we encountered some degeneracies associated with this second clump:
if we did not restrict the value of the velocity dispersion of this second clump to be small, as suggested
by the absence of \textsc{x}-ray emission from that region, we found a solution with an equally
good fit describing this clump
with a high velocity dispersion ($>$ 1\,000 km\,s$^{-1}$) and a large core radius of about
300\,kpc. This solution gives roughly the same projected mass as the solution given in
Table~\ref{tableresults} since for the \textsc{piemd} profile, the mass scales as
$\sigma_0^2 / r_{\mathrm{core}}$. This large core radius is larger than the radius over which the
strong lensing information can be found, thus a clump with such a large radius is an unlikely
solution and we preferred the solution given in Table~\ref{tableresults}.
Given the relatively large number of constraints we have, we find it puzzling to still find
degeneracies in the description of the mass in the nort-east. 

Looking in detail at the modeling of H06 also indicates some problems in that region.
H06 propose two different descriptions of the second clump, one with a non singular isothermal
ellipsoid, which is in agreement with our \textsc{piemd} description, and another one with an elliptical
\textsc{nfw} profile.
Depending on the model they use, the position changes significantly (by about 23$\arcsec$).
Moreover, when using the elliptical \textsc{nfw} profile, a concentration parameter
smaller than 1 is inferred, making this clump very flat, and it is legitimate to ask oneself
about the reality of such a clump, since in numerical simulations, no clumps with a concentration
parameter smaller than 3 do form.

Recent interesting analysis by \citet{saha07} gives more insight on substructures in Abell~1689:
they use two subsets of multiply imaged systems to get two independent mass maps that agree with each other.
Then they substracted off the best \textsc{nfw} fit from these mass maps, and found strong evidence for
substructures independently from the two different data sets, in particular an extended substructure in
the north-east. The substructures they infer are special in the sense that they are characterized by 
extended features much larger than galaxies and more massive than the stellar content, but correlated
with galaxies. They call these extended irregular structures \emph{meso-structures} that appear to be
merging or otherwise dynamically evolving systems.

We find the peak of the total mass distribution to coincide with the center of the \textsc{x}-ray
emission. They both coincide with the \textsc{bcg}.
\citet{zekser} reported an offset between the center of the mass model center and the \textsc{bcg} 
of $\sim 21$ kpc, and H06 also
reported a similar shift of $\sim 16.5$ kpc. It is interesting to
compare to the offset found in \citet{smith05} for a sample of
\textsc{x}-ray luminous clusters whose properties are comparable to
Abell~1689 (i.e. they have a high \textsc{x}-ray luminosity and are
located at $z \sim 0.2$). Considering the irregular clusters in their
sample, they find offsets between the \textsc{x}-ray center and the lensing center of mass
ranging between 10 and 120 kpc.

\subsubsection*{On Galaxy Scales:}
\label{gglensing}

A similar optimization procedure has been performed by \citet{smith05}
for a sample of 6 \textsc{x}-ray luminous clusters, whose properties
are comparable to Abell~1689 (i.e. they have a high \textsc{x}-ray
luminosity and are located at $z \sim 0.2$). They also included the
elliptical galaxies as small scale perturbers, using similar scaling
relations as we use in this work. They find, for a characteristic
luminosity and using a \textsc{piemd} profile, $\sigma_0=180 \pm 20$
km\,s$^{-1}$ and $r_{\mathrm{cut}}=30$\,kpc, in agreement with the
results presented here.

H06 and \citet{aleksi} also find the
elliptical galaxies in Abell~1689 to be truncated, with
$r_{\mathrm{cut}}=60$\,kpc for $\sigma_0=220$ km\,s$^{-1}$ (for a
quite luminous galaxy with a F775W \textsc{ab} magnitude equal to
$\sim$ 16.5), in agreement with the results of this work. \citet{aleksi} investigated
two different possible scaling relations,
$r_{\mathrm{cut}}=r^*_{\mathrm{cut}} \left(\frac{L}{L^*}\right)^{1/2}$
as assumed here, and $r_{\mathrm{cut}}=r^*_{\mathrm{cut}}
\left(\frac{L}{L^*}\right)^{1/4}$, corresponding to $M \propto
L^{-1/4}$. In each case, they found the halos to have a small spatial
extent (compared to field galaxies of equivalent luminosity), and they
were not able to discriminate between these two scaling relations,
since both provide a good fit to the data.

Galaxy dark matter halos in clusters have been probed in earlier work
using galaxy-galaxy lensing techniques
\citep{geigeramas,Priya1,Priya2,Priya3,mypaperII}. 
These studies find cluster galaxies to be significantly
more compact (i.e. with $r_{\mathrm{cut}}<50$ kpc) compared to halos
around field galaxies of equivalent luminosity (for which no clear
edge to the mass distribution has been found so far), in agreement
with tidal stripping (see \citet{mypaperII} for a review of
galaxy-galaxy lensing detections).

\section*{B - New Spectroscopically Confirmed Multiply Imaged Systems}
In this section, we provide some pictures showing the new spectroscopically
confirmed multiply imaged systems we propose in this work.
Fig.~\ref{nicefig} shows the color image of Abell~1689, with the location of
the new multiply imaged systems.
Then the following figures shows each image composing the
different systems. Size of each panel is 27 kpc $\times$ 27 kpc.
For the others multiply imaged systems, see B05 or H06.
Note that the splitting of system 12 from B05 in two different systems
has been first proposed by H06.

\begin{figure}
\epsscale{0.3}
\caption{Systems 12 and 31 at $z=1.83$.}
\label{figimage31}
\end{figure}

\begin{figure}
\epsscale{0.3}
\caption{System 32 at $z=3.0$. 
The image 32.5 in the north west is hardly detected on the color image but we were able to detect
the two little spots on the
F775W image, very close to a cluster member galaxy. Note that without spectroscopy, we would
have been unable to reliably find this counter image.
}
\label{figimage32}
\end{figure}

\begin{figure}
\epsscale{0.3}
\caption{System 33 at $z=4.58$. Image 33.2 corresponds to image 25.2 as identified by B05.}
\label{figimage33}
\end{figure}

\begin{figure}
\epsscale{0.3}
\caption{System 35 at $z=1.52$.} 
\label{figimage35}
\end{figure}

\begin{figure}
\epsscale{0.3}
\caption{System 36 at $z=3.0$.} 
\label{figimage36}
\end{figure}
\begin{figure}
\epsscale{0.3}
\caption{System 40 at $z=2.52$.} 
\label{figimage40}
\end{figure}

\end{document}

%% file: table.tex
\begin{deluxetable}{ccccccccc}
\tabletypesize{\scriptsize}
\tablecaption{Multiply imaged systems considered in this work}

\tablehead{
\colhead{Id} & 
\colhead{\textsc{ra}} & 
\colhead{\textsc{dec}} & 
\colhead{F775W} &  
\colhead{$z_{\mathrm{phot}}$\tablenotemark{a}} & 
\colhead{$z_{\mathrm{model}}$\tablenotemark{b}} & 
\colhead{$z_{\mathrm{spec}}$\tablenotemark{c}} & 
\colhead{\textsc{rms}\tablenotemark{d} (source plane, $\arcsec$)} &
\colhead{\textsc{rms}\tablenotemark{d} (image plane, $\arcsec$)} 
}
\startdata
\noalign{\smallskip}
\noalign{\smallskip}
1.1 &13:11:26.447 &-1:19:56.68 &23.44& 3.03$^{+0.53}_{-0.53}$ & & 3.0 & 0.41& 1.82\\
1.2 &13:11:26.287 &-1:19:59.69 &23.69& 3.04$^{+0.53}_{-0.53}$ & &     & & \\
1.3 &13:11:29.771 &-1:21:07.34 &24.52& 3.27$^{+0.56}_{-0.56}$ & & 3.0 & & \\
1.4 &13:11:33.055 &-1:20:27.31 &24.16& 2.94$^{+0.52}_{-0.52}$ & &  & & \\
1.5 &13:11:31.933 &-1:20:05.69 &24.62& 3.35$^{+0.57}_{-0.57}$ & & & & \\
1.6 &13:11:29.851 &-1:20:38.32 &25.82& 1.06$^{+1.91}_{-0.27}$ & & & & \\
\noalign{\smallskip}
\hline
\noalign{\smallskip}
2.1 &13:11:26.528 &-1:19:55.08 &23.37& 2.62$^{+0.47}_{-0.48}$  & & 2.5 & 0.34 & 1.62 \\
2.2 &13:11:32.961 &-1:20:25.31 &24.18& 2.57$^{+0.47}_{-0.47}$  & & & & \\
2.3 &13:11:31.973 &-1:20:06.89 &24.36& 2.64$^{+0.48}_{-0.48}$  & & & & \\
2.4 &13:11:29.811 &-1:21:05.94 &24.48& 2.36$^{+0.44}_{-0.44}$  & & 2.5& & \\
2.5 &13:11:29.878 &-1:20:39.32 &25.63& 1.59$^{+0.86}_{-0.34}$  && & & \\
\noalign{\smallskip}
\hline
\noalign{\smallskip}
3.1 &13:11:32.043 &-1:20:27.41 &26.65& 5.48 $^{+0.85}_{-0.85}$ & 3.25 $\pm$1.47& & 0.32 & 0.55\\
3.2 &13:11:32.174 &-1:20:33.31 &26.85& 5.45 $^{+0.85}_{-0.85}$ & & & & \\
3.3 &13:11:31.680 &-1:20:55.88 &   &  &  & & &\\
\noalign{\smallskip}
\hline
\noalign{\smallskip}
4.1 &13:11:32.174 &-1:20:57.33 &24.62& 1.06$^{+0.27}_{-0.27}$ &  & 1.1 & 0.42 & 1.29\\
4.2 &13:11:30.518 &-1:21:11.94 &23.91& 1.32$^{+0.30}_{-0.31}$ &  & & & \\
4.3 &13:11:30.759 &-1:20:07.89 &25.12& 1.47$^{+0.32}_{-0.33}$ &  & & & \\
4.4 &13:11:26.287 &-1:20:35.11 &24.67& 1.33$^{+0.31}_{-0.31}$ &  & & & \\
4.5 &13:11:29.838 &-1:20:29.11 &  &   &  & & & \\
\noalign{\smallskip}
\hline
\noalign{\smallskip}
5.1 &13:11:29.064 &-1:20:48.33 &24.42& 3.29$^{+0.56}_{-0.56}$ & & 2.6 &0.72 &2.11 \\
5.2 &13:11:29.224 &-1:20:43.92 &24.92& 3.16$^{+0.55}_{-0.55}$ & & & & \\
5.3 &13:11:34.109 &-1:20:20.90 &25.26& 2.15$^{+0.67}_{-0.41}$ & & 2.6 & & \\
\noalign{\smallskip}
\hline
\noalign{\smallskip}
6.1 &13:11:30.759 &-1:19:37.87 &23.61& 1.22$^{+0.29}_{-0.29}$ & & 1.1 & 0.54 & 3.35 \\
6.2 &13:11:33.335 &-1:20:12.10 &23.85& 1.31$^{+0.30}_{-0.30}$ & & 1.1 & & \\
6.3 &13:11:32.734 &-1:19:54.28 &23.02& 0.94$^{+0.26}_{-0.25}$ & & & & \\
6.4 &13:11:32.481 &-1:19:58.69 &24.00& 1.09$^{+0.27}_{-0.27}$ & & & & \\
\noalign{\smallskip}
\hline
\noalign{\smallskip}
7.1 &13:11:25.446 &-1:20:51.53 &23.31& 4.92$^{+0.78}_{-0.78}$ && 4.8 & 0.50 & 1.07\\
7.2 &13:11:30.679 &-1:20:13.70 &24.19& 5.20$^{+0.81}_{-0.81}$ & &4.8 & & \\
7.3 &13:11:29.824 &-1:20:24.71 &28.23& 0.77$^{+4.01}_{-0.23}$ & & & & \\
\noalign{\smallskip}
\hline
\noalign{\smallskip}
8.1 &13:11:32.300 &-1:20:50.78 &24.54& 2.63$^{+0.48}_{-0.48}$ & 2.30$\pm$0.21 & & 0.55 & 2.49\\
8.2 &13:11:31.393 &-1:21:05.59 &24.31& 2.77$^{+0.50}_{-0.50}$ & & & & \\
8.3 &13:11:31.500 &-1:20:13.95 &25.71& 2.75$^{+0.49}_{-0.89}$ & & & & \\
8.4 &13:11:25.520 &-1:20:20.15 &23.97& 0.70$^{+0.22}_{-0.22}$ & & & & \\
8.5 &13:11:30.325 &-1:20:30.46 &27.42& 0.77$^{+2.53}_{-0.23}$ & & & & \\
\noalign{\smallskip}
\hline
\noalign{\smallskip}
9.1 &13:11:30.298 &-1:19:48.33 &25.92& 4.97$^{+0.78}_{-0.78}$ &2.69$\pm$0.27 & & 0.36 & 1.60 \\
9.2 &13:11:33.515 &-1:20:50.38 &27.53& 1.06$^{+0.27}_{-0.27}$ & & & & \\
9.3 &13:11:28.723 &-1:21:15.60 &25.72& 5.16$^{+0.81}_{-0.81}$ & & & & \\
9.4 &13:11:26.267 &-1:20:26.56 &27.15& 5.17$^{+0.81}_{-0.81}$ & & & & \\
\noalign{\smallskip}
\hline
\noalign{\smallskip}
10.1 &13:11:33.976 &-1:20:50.93 &23.66& 1.75$^{+0.74}_{-0.36}$ & & 1.8 & 0.19 & 0.43\\
10.2 &13:11:28.049 &-1:20:12.30 &23.30& 1.54$^{+0.33}_{-0.33}$ & & & & \\
10.3 &13:11:29.317 &-1:20:27.71 &25.15& 2.57$^{+0.47}_{-0.63}$ & & & & \\
\noalign{\smallskip}
\hline
\noalign{\smallskip}
11.1 &13:11:33.333 &-1:21:06.74 &24.03& 2.91$^{+0.51}_{-0.51}$ & &2.5 & 0.22 & 0.60\\
11.2 &13:11:29.064 &-1:20:01.09 &23.36& 2.87$^{+0.51}_{-0.51}$ & & & & \\
11.3 &13:11:29.491 &-1:20:26.31 &26.78& 1.58$^{0.73}_{-0.52}$ & & & & \\
\noalign{\smallskip}
\hline
\noalign{\smallskip}
12.2 &13:11:27.355 &-1:20:54.73 &24.30& 1.99$^{+0.39}_{-0.39}$ & & 1.8 & 0.04 & 1.54\\
12.3 &13:11:27.208 &-1:20:51.53 &23.97& 1.99$^{+0.39}_{-0.39}$ & & & & \\
\noalign{\smallskip}
\hline
\noalign{\smallskip}
13.1 &13:11:32.821 &-1:19:24.11 &24.10& 1.02$^{+0.28}_{-0.27}$ & 1.5$\pm$0.5& & 0.38 & 4.72\\
13.2 &13:11:32.994 &-1:19:25.51 &23.68& 0.72$^{+0.23}_{-0.23}$ & & & & \\
13.3 &13:11:33.408 &-1:19:31.11 &24.12& 1.10$^{+0.28}_{-0.28}$ & & & & \\
\noalign{\smallskip}
\hline
\noalign{\smallskip}
14.1 &13:11:29.024 &-1:21:41.77 &24.98& 3.37$^{+0.57}_{-0.82}$ & & 3.4 & 0.12 & 4.02\\
14.2 &13:11:29.454 &-1:21:42.67 &25.99& 3.64$^{+0.61}_{-0.61}$ & & & & \\
\noalign{\smallskip}
\hline
\noalign{\smallskip}
15.1 &13:11:28.083 &-1:20:14.75 &25.69& 1.99$^{+0.39}_{-0.39}$ &  &1.8 &  0.15 & 0.38\\
15.2 &13:11:34.076 &-1:20:51.38 &25.79& 2.00$^{+0.39}_{-0.39}$ & & & & \\
15.3 &13:11:29.244 &-1:20:27.36 &27.16& 1.97$^{+0.39}_{-0.43}$ & & & &  \\
\noalign{\smallskip}
\hline
\noalign{\smallskip}
16.1 &13:11:27.989 &-1:20:24.96 &23.68& 1.81$^{+0.37}_{-0.37}$ &1.76$\pm$0.16& & 0.47 & 2.78\\
16.2 &13:11:28.923 &-1:20:28.36 &25.06& 2.26$^{+0.43}_{-0.43}$ && &  & \\
16.3 &13:11:34.396 &-1:20:46.37 &25.35& 1.80$^{+0.71}_{-0.37}$ && &  & \\
\noalign{\smallskip}
\hline
\noalign{\smallskip}
17.1 &13:11:30.655 &-1:20:24.76 &24.19& 2.74$^{+0.49}_{-0.49}$ & & & 0.55 & 1.13 \\
17.2 &13:11:30.388 &-1:20:27.56 &25.11& 2.02$^{+0.40}_{-0.40}$ & &  & & \\
17.3 &13:11:24.983 &-1:20:41.37 &24.31& 2.25$^{+0.43}_{-0.43}$ & & 2.6 & & \\
\noalign{\smallskip}
\hline
\noalign{\smallskip}
18.1 &13:11:28.243 &-1:20:09.34 &25.00& 2.56$^{+0.47}_{-0.47}$ &  & 1.8 & 0.07 & 0.19\\
18.2 &13:11:33.809 &-1:20:54.58 &  &   &  &  &  & \\
18.3 &13:11:29.364 &-1:20:27.16 &26.78& 1.58$^{+0.73}_{-0.52}$ && & & \\
\noalign{\smallskip}
\hline
\noalign{\smallskip}
19.1 &13:11:31.633 &-1:20:22.56 &24.83& 1.72$^{+0.36}_{-0.36}$ & & & 0.78 & 1.75 \\
19.2 &13:11:25.253 &-1:20:19.75 &25.23& 2.74$^{+0.49}_{-0.49}$ & &  & & \\
19.3 &13:11:31.953 &-1:20:59.18 &25.88& 1.57$^{+0.34}_{-0.34}$ & &  & & \\
19.4 &13:11:32.033 &-1:20:57.38 &24.76& 2.58$^{+0.47}_{-0.47}$ & & 2.6 & & \\
19.5 &13:11:30.218 &-1:20:33.76 &27.94& 4.54$^{+1.66}_{-0.73}$ & & &  & \\
\noalign{\smallskip}
\hline
\noalign{\smallskip}
21.1 &13:11:31.019 &-1:20:45.77 &25.15& 1.79$^{+0.37}_{-0.37}$ &1.91$\pm$0.19 & & 0.35 & 1.76\\
21.2 &13:11:30.792 &-1:20:44.57 &26.70& 1.59$^{+0.34}_{-0.34}$ && & & \\
21.3 &13:11:25.253 &-1:20:10.95 &25.59& 1.78$^{+0.36}_{-0.36}$ &&  & & \\
\noalign{\smallskip}
\hline
\noalign{\smallskip}
22.1 &13:11:29.694 &-1:20:08.54 &23.66& 1.99$^{+0.39}_{-0.39}$ && 1.7 & 0.25 &0.69 \\
22.2 &13:11:29.614 &-1:20:23.56 &25.49& 1.99$^{+0.59}_{-0.39}$ && & & \\
22.3 &13:11:32.404 &-1:21:16.00 &23.27& 1.96$^{+0.39}_{-0.39}$ &&  & & \\
\noalign{\smallskip}
\hline
\noalign{\smallskip}
23.1 &13:11:29.524 &-1:20:09.74 &24.63& 2.03$^{+0.40}_{-0.40}$ &1.78$\pm$0.20&& 0.14 &0.44 \\
23.2 &13:11:29.551 &-1:20:22.76 &25.99& 1.99$^{+0.62}_{-0.39}$ & &  & & \\
23.3 &13:11:32.661 &-1:21:15.20 &24.72& 2.00$^{+0.39}_{-0.39}$ && & & \\
\noalign{\smallskip} 
\hline
\noalign{\smallskip}
24.1 &13:11:29.187 &-1:20:55.98 &25.24& 2.63$^{+0.48}_{-0.48}$ && & 0.51 & 4.52\\
24.2 &13:11:32.057 &-1:19:50.33 &24.80& 2.50$^{+0.46}_{-0.46}$ &&  & &  \\
24.3 &13:11:30.295 &-1:19:33.92 &24.33& 2.43$^{+0.45}_{-0.45}$ && 2.6 & & \\
24.4 &13:11:33.712 &-1:20:19.75 &24.99& 2.81$^{+0.50}_{-0.69}$ &&  & & \\
\noalign{\smallskip}
\hline
\noalign{\smallskip}
26.1 &13:11:25.146 &-1:20:32.36 &24.67& 1.08$^{+0.39}_{-0.27}$ &2.33$\pm$0.29&  & 0.66 & 2.15\\
26.2 &13:11:31.326 &-1:20:25.16 &25.39& 1.04$^{+0.27}_{-0.27}$ & &  & & \\
26.3 &13:11:30.245 &-1:20:32.36 &27.42& 0.77$^{+2.53}_{-0.23}$ & &  & & \\
\noalign{\smallskip}
\hline
\noalign{\smallskip}
27.1 &13:11:25.174 &-1:20:33.11 &25.22& 1.81$^{+0.37}_{-0.37}$ &2.38$\pm$0.28& & 0.78 & 2.56 \\
27.2 &13:11:31.366 &-1:20:24.56 &26.19& 1.58$^{+0.48}_{-0.34}$ && & & \\
27.3 &13:11:30.191 &-1:20:32.76 &29.82& 4.55$^{+1.63}_{-0.73}$ && & & \\
\noalign{\smallskip}
\hline
\noalign{\smallskip}
28.1 &13:11:28.298 &-1:20:10.91 &27.20& 1.17$^{+4.29}_{-0.29}$ &2.93$\pm$0.48& & 0.57 & 1.71 \\
28.2 &13:11:34.260 &-1:21:00.01 &26.47& 2.00$^{+0.43}_{-1.23}$ & &  & & \\
\noalign{\smallskip}
\hline
\noalign{\smallskip}
29.1 &13:11:29.240 &-1:20:57.78 &25.97& 2.47$^{+0.57}_{-0.46}$ && & 0.41 & 2.65 \\
29.2 &13:11:30.028 &-1:19:33.92 &25.00& 3.40$^{+0.58}_{-0.58}$ &&  & & \\
29.3 &13:11:32.164 &-1:19:52.53 &24.80& 2.50$^{+0.46}_{-0.46}$ && & & \\
29.4 &13:11:33.618 &-1:20:20.55 &25.84& 3.35$^{+0.57}_{-0.57}$ && 2.5 & & \\ 
\noalign{\smallskip}
\hline
\noalign{\smallskip}
30.1 &13:11:32.424 &-1:19:19.50 &25.91& 4.49$^{+0.72}_{-0.72}$ & & &0.61 & 11.20\\
30.2 &13:11:33.188 &-1:19:25.81 &25.80& 3.23$^{+0.56}_{-0.76}$ & & & & \\
30.3 &13:11:33.662 &-1:19:32.51 &25.73& 3.30$^{+0.56}_{-0.56}$ & & 3.0 & & \\
\noalign{\smallskip}
\hline
\noalign{\smallskip}
31.1 & 13:11:30.362 & -01:19:51.13 & 23.77 & &  & 1.8 &  0.34 & 1.1\\
31.2 & 13:11:33.271 &-01:20:44.37  &  &  &  & & & \\
31.3 &13:11:28.960 &-01:21:10.19  & 25.37& &  &  & & \\
31.4 & 13:11:26.504&-01:20:21.75 & 24.80 & &  & & & \\
\noalign{\smallskip}
\hline
\noalign{\smallskip}
32.1 & 13:11:32.190 & -01:20:03.34 & & & & & 0.18 & 0.75\\
32.2 & 13:11:33.218 & -01:20:20.75 & & & & 3.0 & & \\
32.3 & 13:11:29.587 & -01:21:02.49 & & & & & & \\
32.4 & 13:11:29.801 & -01:20:43.17 & & & & & & \\
32.5 & 13:11:26.600 & -01:19:57.60 & & & & 3.0 & & \\
\noalign{\smallskip}
\hline
\noalign{\smallskip}
33.1 &13:11:28.440 &-01:21:00.39 &23.66&  &&4.5 &0.51 & 1.56\\
33.2 &13:11:34.646 &-01:20:33.56 &26.62& && & & \\
\noalign{\smallskip}
\hline
\noalign{\smallskip}
35.1 & 13:11:28.493 & -01:20:59.38 & & & & 1.9 & 0.48 & 1.52\\
35.2 & 13:11:33.952 & -01:20:33.36 & & & &  & & \\
35.3 & 13:11:29.427 & -01:20:33.76 & & & &  & & \\
\noalign{\smallskip}
\hline
\noalign{\smallskip}
36.1 & 13:11:31.563 & -01:19:45.73 &  & & & 3.0 & 0.08 & 1.15\\
36.2 & 13:11:31.683 & -01:19:47.13 & &  & &  &  & \\
\noalign{\smallskip}
\hline
\noalign{\smallskip}
40.1 & 13:11:30.269 & -01:20:11.43 &  & & & 2.5 & 0.39 & 0.99\\
40.2 & 13:11:26.174 & -01:21:02.54 & &  & &  &  & \\
\noalign{\smallskip}
\hline
\noalign{\smallskip}

\enddata
\tablenotetext{a}{Photometric redshift, from B05}
\tablenotetext{b}{Redshift estimation infered from the mass model when all spectroscopically confirmed multiply imaged
system have been included in the optimization. Errors bars quote 3$\sigma$ confidence level}
\tablenotetext{c}{Spectroscopic redshift, see Paper I}
\tablenotetext{d }{The mean scatter is given for the whole system, not for each individual image
composing a system}
\label{multipletable}
\end{deluxetable}